\documentclass[twocolumn,aps,prd,reprint,nofootinbib,url = false,longbibliography,superscriptaddress,en]{revtex4-2}
\usepackage[utf8]{inputenc}
\usepackage[english]{babel}
\usepackage{amssymb}
\usepackage{amsmath}
\usepackage{comment}
\usepackage{lipsum}
\usepackage{bm}
\usepackage{booktabs}

\usepackage{mhchem}

\usepackage{mathtools}
\usepackage[final]{graphicx}

\usepackage[labelformat=simple,subrefformat=parens,labelformat=parens]{subfig}
\usepackage{floatrow}
\usepackage{caption}
\newcommand{\RN}[1]{%
  \textup{\uppercase\expandafter{\romannumeral#1}}%
}

\newcommand{\com}[1]{\textcolor{red}{ #1}}

\usepackage[usenames, dvipsnames]{color}

\definecolor{myred1}{RGB}{255, 0, 0}
\definecolor{myyellow1}{RGB}{255, 255, 219}
\definecolor{mygreen1}{RGB}{0, 255, 0}
\definecolor{mygreen2}{RGB}{0, 126, 0}
\definecolor{myblue1}{RGB}{0, 0, 255}
\usepackage{physics}

\begin{document}
\title{
Excitability
and oscillations 
of active droplets
} 

\author{Ivar S. Haugerud}
\affiliation{
 Faculty of Mathematics, Natural Sciences, and Materials Engineering: Institute of Physics, University of Augsburg, Universit\"atsstra\ss e~1, 86159 Augsburg, Germany
}

\author{Hidde D. Vuijk}
\affiliation{
 Faculty of Mathematics, Natural Sciences, and Materials Engineering: Institute of Physics, University of Augsburg, Universit\"atsstra\ss e~1, 86159 Augsburg, Germany
}

\author{Job Boekhoven}

\affiliation{School of Natural Sciences, Department of Bioscience, Technical University of Munich, Lichtenbergstraße 4, 85748 Garching, Germany}

\author{Christoph A. Weber}

\affiliation{
 Faculty of Mathematics, Natural Sciences, and Materials Engineering: Institute of Physics, University of Augsburg, Universit\"atsstra\ss e~1, 86159 Augsburg, Germany
}
\date{\today}

\begin{abstract}
In living cells, cycles of 
formation and dissolution of liquid droplets can mediate biological functions such as DNA repair. 
However, the minimal physicochemical prerequisite for such droplet oscillations remains elusive. 
Here, we present a simple model composed of only two independent chemical components with their diffusive and chemical fluxes
governed by non-equilibrium thermodynamics.
There is turnover of fuel that maintains a chemical reaction away from equilibrium, leading to active droplets. 
We find that a single active droplet undergoes a pitchfork-bifurcation in the droplet volume upon increasing the fueling strength. 
Strikingly, the active droplet becomes excitable upon adding a further chemical reaction.  
For sufficient fueling, the system undergoes self-sustained oscillations cycling between droplet formation and dissolution.
The minimal nature of our model suggests  
self-sustained active droplets as functional modules for \textit{de novo} life. 
\end{abstract}
\maketitle

\textit{Introduction:}
Oscillations of concentrations are fundamental to numerous biological processes, including the circadian rhythm~\cite{zehringPelementTransformationPeriod1984,bargielloRestorationCircadianBehavioural1984} and the cell cycle~\cite{dunlapMolecularBasesCircadian1999,goldbeterBiochemicalOscillationsCellular1996}. Extensive experimental and theoretical work has revealed the underlying mechanism of oscillating chemical reactions, such as the Briggs–Rauscher reaction~\cite{brayPERIODICREACTIONHOMOGENEOUS1921,zhabotinskyHistoryChemicalOscillations1991}, the Belousov–Zhabotinsky reaction~\cite{pechenkinBelousovHisReaction2009,belousovPeriodicReactionIts1958}, and the Brusselator~\cite{prigogineSymmetryBreakingInstabilitiesDissipative1967,prigogineSymmetryBreakingInstabilities1968}. The oscillations emerge as a consequence of non-linearities in the chemical kinetic equations giving rise to an unstable region in phase space through a Hopf-bifurcation~\cite{marsdenHopfBifurcationIts2012,caraballoStochasticpitchforkBifurcation2001}. Such non-linearities can further lead to excitable concentration fronts and pulses that propagate, as well as stationary Turing patterns~\cite{cross1993pattern,turingChemicalBasisMorphogenesis1997,epsteinNonlinearChemicalDynamics1996,novakDesignPrinciplesBiochemical2008, halatekRethinkingPatternFormation2018}. 

An alternative mechanism for spatial patterns in chemically active systems is through phase separation of liquid condensates~\cite{sodingMechanismsActiveRegulation2020,folkmannRegulationBiomolecularCondensates2021,hymanLiquidLiquidPhaseSeparation2014c,bananiBiomolecularCondensatesOrganizers2017}. The patterns not necessarily originate from non-linearities in the chemical kinetics but from thermodynamic interactions between the components~\cite{weberPhysicsActiveEmulsions2019d,luo2023influence,menou2023physical}. When the chemical reactions break detailed balance of the rates, the system is maintained away from equilibrium, making the droplets active~\cite{riess2020design,zwickerChemicallyActiveDroplets2024,weberPhysicsActiveEmulsions2019d}. Active droplets can behave fundamentally different from their equilibrium counterpart, exhibiting properties such as droplet  division~\cite{zwickerGrowthDivisionActive2017b},
accelerated ripening~\cite{tena2021accelerated},
size control~\cite{wurtzChemicalReactionControlledPhaseSeparated2018,zwickerGrowthDivisionActive2017b,zwickerSuppressionOstwaldRipening2015}, 
shell-formation~\cite{bergmannLiquidSphericalShells2023d}, and various other spatial morphologies and patterns~\cite{bartolucciControllingCompositionCoexisting2021d,bauermannCriticalTransitionIntensive2024}.

Oscillating chemical reactions rely on the precise spatial and temporal regulation of chemical processes -- a property that can be provided by biomolecular condensates~\cite{wurtzChemicalReactionControlledPhaseSeparated2018,hymanLiquidLiquidPhaseSeparation2014c,bananiBiomolecularCondensatesOrganizers2017}. A key question is whether liquid condensates could not only regulate the chemical processes to enhance the stability of oscillating chemical reactions~\cite{smokersPhaseSeparatedDropletsCan2025}, but create the oscillations themselves. Cyclic reaction networks coupled to the liquid condensate could lead to oscillations in concentrations and even of the droplet. Such a mechanism is of biological relevance, as cycles of formation and dissolution of active biomolecular condensates  have been shown to facilitate DNA repair~\cite{hsiehPlausibleRobustBiological2024,heltbergEnhancedDNARepair2022,heltbergCoupledOscillatorCooperativity2023}. 

\begin{figure}
    \centering
    \makebox[\textwidth][c]{{\includegraphics[width=\textwidth]{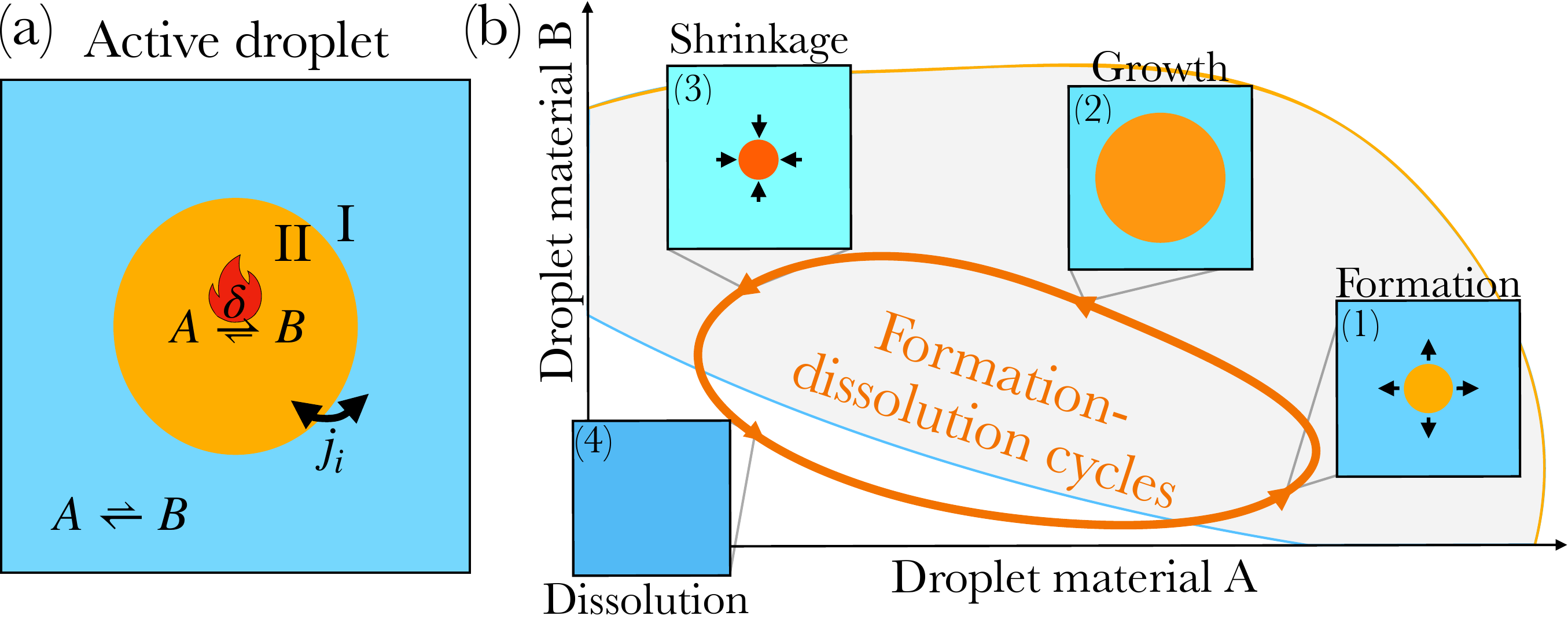}}}
     \caption{
     \textbf{Formation and dissolution of active droplets:} (a) Chemically active droplets break detailed balance of the rates through the turnover of a fuel of strength $\delta$. This turnover drives the reaction of component $A$ to $B$ in the dense phase (II), while in the dilute phase (I), there is no fuel, and the pathway from $B$ to $A$ dominates (Eq.~\eqref{eq:r_1_pathway}). We find that for a large enough fuel strength $\delta$, a pitchfork bifurcation in droplet volumes occurs.
     (b) Upon adding a further chemical reaction (Eq.~\eqref{eq:r_2_pathway}) that breaks the conservation of $A$ and $B$, makes the active droplet excitable and capable of self-sustained cycles of formation (1) and dissolution (4). 
     }
     \label{fig:1}
\end{figure}

Our work presents a minimal theoretical model to achieve self-sustained oscillation of condensates; see Fig.~\ref{fig:1}. The oscillations do not emerge due to a complex non-linear reaction network but through the combination of minimalistic chemistry away from equilibrium and the thermodynamics of phase separation, leading to an active droplet. We show that a pitchfork bifurcation in the droplet volume emerges as active droplets are driven far away from equilibrium, making them bistable. With the addition of a second reaction that breaks the conservation law of the first chemical reaction, the droplets behave as an excitable medium, exhibiting both wave propagation with a refractory period and emergent self-sustained oscillations.

\begin{figure*}
    \centering
     \makebox[\textwidth][c]{
\includegraphics[width=\textwidth]{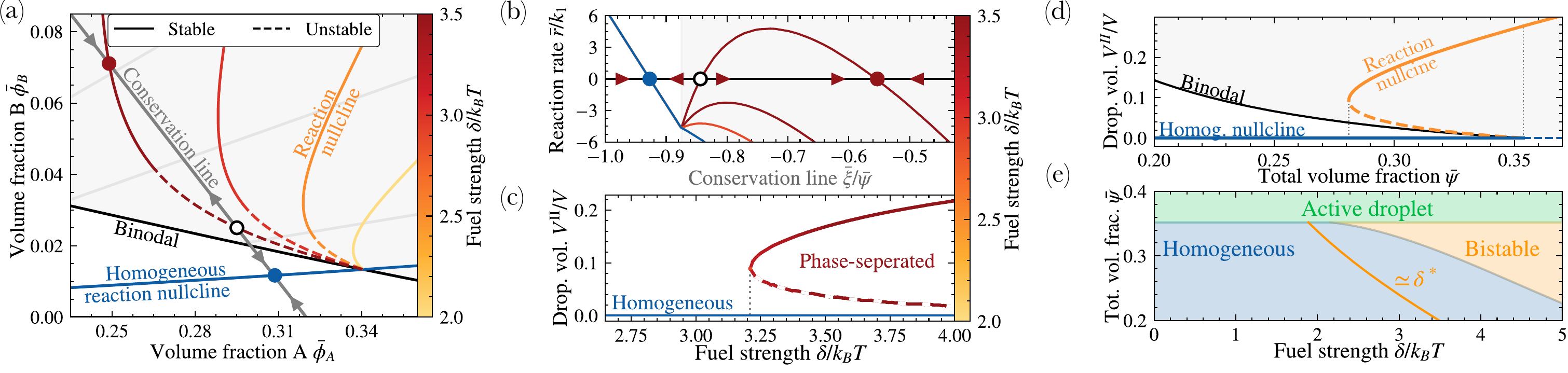}}
     \caption{
     \textbf{Pitchfork bifurcation to an active droplet:}
     (a) Upon increasing the fuel strength $\delta$, the reaction nullcline (Eq.~\eqref{eq:NESS}) for reaction $r_1$ (Eq.~\eqref{eq:r_1_pathway}) in the phase-separated domain (gray shaded) bends, leading to three non-equilibrium steady states (NESS) for a fixed conservation line $\bar{\psi}=\bar{\phi}_A+\bar{\phi}_B$; closed symbols are stable and open symbols unstable fixed points. 
     (b) The net reaction rate along the conservation line only has multiple steady states for sufficiently large fuel strength $\delta/(k_BT)$, where the center point is unstable.
     (c) The multiple steady points arise from a pitchfork bifurcation in droplet volume upon a critical fuel strength $\delta^*$ (vertical dotted line). Above this threshold value  $\delta^*$, there are three NESS' where two are locally stable (bistability).
     (d) 
      At fixed fuel strength,
     The occurrence of the bifurcation depends on the value of quantity $\bar{\psi}$ conserved by reaction~\eqref{eq:r_1_pathway}. There is a range of  $\bar{\psi}$-values (vertical dotted lines), where the system is bistable. 
     (e) 
     Bistability is only possible within a range of $\bar{\psi}$ values that grow with fuel strength $\delta$. The analytical approximation for the critical fuel strength $\delta^*$ from Eq.~\eqref{eq:full_delta*} agrees well for fuel strength $\delta$ close to the onset of bistability.}
     \label{fig:2}
\end{figure*}

\textit{Theory of active droplets:}
The evolution of volume fractions $\phi_i(\bm{x}, t)$ in space $\bm{x}$ and time $t$ are governed by diffusive fluxes $\bm{j}_i$ and reaction rates $r_i$ of component $i$ with the corresponding continuous equations~\cite{julicherDropletPhysicsIntracellular2024, bauermannEnergyMatterSupply2022d,weberPhysicsActiveEmulsions2019d}:
\begin{equation}   \partial_t\phi_i(\bm{x}, t) =  -\nabla \cdot \bm{j}_i + r_i\,  .\label{eq:phi_dot}
\end{equation}
We investigate a minimalistic ternary system $(i=A,B,C)$ initially limited to a single reversible uni-molecular chemical reaction between $A$ and $B$,
\begin{equation}
r_1: \quad \ce{A <=>>[$k_1\exp{\tilde{\delta}/(k_BT)}$][$k_1$]B} \, ,\label{eq:r_1_pathway}
\end{equation}
where the net reaction rate $r_B=-r_A= r_1$ and $k_1$ is the reaction rate coefficient. The reaction rates are governed by the chemical potentials $\mu_i$~\cite{adame-aranaLiquidPhaseSeparation2020a,bauermannChemicalKineticsMass2022d} 
(see Appendix~\ref{seq:free_energy_and_chempot}),
\begin{equation}
    r_1 = k_1\left[ \exp{\frac{\mu_A + \tilde{\delta}(\phi_A, \phi_B)}{k_BT}} - \exp{\frac{\mu_B}{k_BT}} \right]\, , \label{eq:r}
\end{equation}
where $k_BT$ is the thermal energy. Moreover, $\tilde{\delta}(\phi_A,\phi_B)$ captures the effects of a composition-dependent fuel~\cite{boekhovenDissipativeSelfAssemblyMolecular2010,boekhovenTransientAssemblyActive2015} that enhances the forward reaction pathway, where $\exp\{ \tilde \delta/(k_BT)\}$ can be understood as a chemostatted fuel activity; see Appendix~\ref{sec:fuel} for further discussions. 
A non-vanishing $\tilde{\delta}(\phi_A,\phi_B)$ breaks detailed balance of the rates, $r_1^\rightharpoonup/r_1^\leftharpoondown = \exp\{(\Delta\mu + \tilde{\delta})/(k_BT)\}$, where $\Delta\mu=\mu_A-\mu_B$ characterizes how far the system is away from chemical equilibrium~\cite{julicherModelingMolecularMotors1997b,weberPhysicsActiveEmulsions2019d}.

According to Eq.~\eqref{eq:phi_dot}, the volume fraction fields $\phi_i$ further evolve due to chemical potential gradients in space, leading to diffusive fluxes. In linear response, 
diffusive fluxes read $\bm{j}_i = -\Lambda_i\nabla \mu_i$,
where $\Lambda_i$ are the mobility coefficients~\cite{de2013non}.
For sufficiently strong interactions between the components, the system can phase-separate, creating  stable inhomogeneous volume-fraction profiles.

The dynamic Eqs.~\eqref{eq:phi_dot} governing phase separation simplify~\cite{bauermannChemicalKineticsMass2022d} when the diffusive exchange rate between phases is much faster than the reaction rate coefficient, $k_1 \ll \Lambda_ik_BT/L^2$, where $L$ is the system size and $\sqrt{\Lambda_ik_BT/k_1}$ is the corresponding reaction-diffusion length scale.
In this limit, the two phases $(\RN{1},\,\RN{2})$ satisfy the conditions of phase equilibrium at all times:
\begin{equation}
    \mu_i^\RN{1} = \mu_i^\RN{2}\,. \label{eq:phase_coex}
\end{equation} Furthermore, each phase is homogeneous in composition with different volume fractions in each phase $(\phi_i^\RN{1},\phi_i^\RN{2})$ with corresponding phase volumes $V^\RN{1}$ and $V^{\RN{2}}=V-V^\RN{1}$. This leads to the simplified  dynamic equations at phase equilibrium~\cite{bauermannChemicalKineticsMass2022d}:
\begin{subequations}
\label{eq:kinetic_evolve}
\begin{align}
    \dv{\phi_i^{\RN{1}/\RN{2}}}{t} &= r_i^{\RN{1}/\RN{2}} - j^{\RN{1}/\RN{2}}_i - \phi_i^{\RN{1}/\RN{2}}\frac{\dot{V}^{\RN{1}/\RN{2}}}{V^{\RN{1}/\RN{2}}} \, , \label{eq:phi_dot_fast_diff}\\
    \frac{\dot{V}^{\RN{1}/\RN{2}}}{V^{\RN{1}/\RN{2}}} &= \sum_{k=A,B,C} \left( r_k^{\RN{1}/\RN{2}} - j_k^{\RN{1}/\RN{2}} \right)\, , \label{eq:vol_change}
\end{align} 
\end{subequations}
where $i=A,B$, and the diffusive flux $j_i^{\RN{1}/\RN{2}}$ is set to ensure the conditions of phase coexistence (Eq.~\eqref{eq:phase_coex}), see Appendix~\ref{seq:SJ_method} for a derivation. Phase-averaged quantities are denoted by a bar, which, for an arbitrary quantity $X$ at phase equilibrium, is
\begin{equation}
    \bar{X} \equiv \frac{V^\RN{1}X^\RN{1} + V^\RN{2}X^\RN{2}}{V^\RN{1} + V^\RN{2}}\,. \label{eq:phase_avg}
\end{equation}
For a ternary system with chemical reaction 1 (Eq.~\eqref{eq:r_1_pathway}), a natural choice of quantities is the phase-averaged total volume fraction $\bar{\psi}$, which is conserved during reaction 1, and the phase-averaged reaction extent $\bar{\xi}$,
\begin{equation}
    \bar{\psi} \equiv \frac{\bar{\phi}_A+\bar{\phi}_B}{2}\,,\qquad \qquad\quad  \bar{\xi} \equiv \frac{\bar{\phi}_B-\bar{\phi}_A}{2}\,.\label{eq:tot_vol_frac}
\end{equation}
With this choice of quantities, Eqs.~\eqref{eq:kinetic_evolve} become
\begin{equation}
\label{eq:dt_xi_bar}
 \text{d}_t\bar{\psi}=0\,,\quad\qquad\qquad \text{d}_t\bar{\xi}=\bar{r}_1   \,.
\end{equation}
Steady states, which corresponds to $\bar{r}_1=0$, defines the reaction nullcline of $r_1$:
\begin{equation}
    \label{eq:nullcline_condtion}
 V^\RN{1}\,r^\RN{1}_1 = V^\RN{2}\,r^\RN{2}_1\, .
\end{equation}

For reactions obeying detailed balance of the rates ($\tilde \delta=0$), the nullcline is the tie-line of the binodal, making it independent of the phase volumes ($V^\RN{1}$ and $V^\RN{2}$). For chemical reactions that break the detailed balance of the rates, the nullcline for phase-separated systems depends on the phase volumes. To maintain the system away from equilibrium, 
we consider 
\begin{equation}
\label{eq:delta_tilde}
\tilde{\delta}(\psi) = \frac{\delta}{\pi}\left[\frac{\pi}{2} + \tanh{\left(\frac{\psi-\psi_{1/2}}{\kappa}\right)}\right] 
\, ,
\end{equation}
where 
$\delta\geq 0$ is the fuel strength that enhances turnover from $A$ to $B$. For $\delta=0$, the system can settle to thermodynamic equilibrium. 
The parameter $\kappa\ll 1$ 
sets the range of $\psi$-values around $\psi_{1/2}$ where the transition from non-fueled to the fueled case occurs.
For $\psi_{1/2}$ inside the binodal, the dilute phase is  approximately unfueled, while the dense phase is   fueled  ($\tilde{\delta}^\RN{1}\simeq 0$, $\tilde{\delta}^\RN{2}\simeq \delta$), resulting in a volume-dependent reaction nullcline when phase-separated.
The phase-dependent fueling gives rise to non-equilibrium steady states (NESS) with continuous chemical turnover in both phases, balanced by diffusive fluxes across the phase boundary, see Fig.~\ref{fig:1}(a). For the fueling described by Eq.~\eqref{eq:delta_tilde}, the component $B$ is produced in the dense phase (large $\psi$) and diffuses to the dilute phase (small $\psi$), reacting to $A$ before diffusing back to the droplet. 
Eq.~\eqref{eq:nullcline_condtion}
dictates a corresponding steady state droplet volume
\begin{equation}
    \frac{V^\RN{2}}{V} = \frac{k_1^\RN{1}\left[e^{\frac{\Delta\mu+\delta^\RN{1}}{k_BT}} - 1 \right]}{k_1^\RN{2}\left[e^{\frac{\Delta\mu+\delta^\RN{2}}{k_BT}} - 1\right] -  k_1^\RN{1}\left[e^{\frac{\Delta\mu+\delta^\RN{1}}{k_BT}} - 1 \right] } \, ,  \label{eq:NESS}
\end{equation}
where $\delta^\text{I/II}=\tilde{\delta}(\psi^\text{I/II})$ and $k_1^\text{I/II}$ denote the phase dependent reaction rate coefficients of reaction 1 (Eq.~\eqref{eq:r_1_pathway}). 
Unless stated otherwise, we solve Eqs.~\eqref{eq:kinetic_evolve} numerically. 
For information about the parameters used and numerical implementation, see Appendix~\ref{seq:parameters_used}.

\textit{Pitchfork bifurcation to an active droplet:} 
Without fuel ($\delta=0$), the steady state of the system is homogeneous, corresponding to thermodynamic equilibrium.
Upon increasing the fuel strength
$\delta$, active droplets emerge as non-equilibrium steady states that are locally stable.
Their emergence requires a large enough fueling, bending the reaction nullcline strong enough (Fig.~\ref{fig:2}(a,b)) and thereby creating multiple intersections with the same conserved line of constant total volume fraction $\bar{\psi}$.
We note that the homogeneous steady state remains stable as $\delta$ is increased and a locally stable active droplet is created. Thus, the system becomes bistable with two locally stable steady states. 

Bistability between an active droplet state and a homogeneous state emerges  through a pitchfork bifurcation at which the steady droplet volume $V^\RN{2}$ jumps from zero to a finite value;
see Fig.~\ref{fig:2}(c). 
The bifurcation occurs at a critical fuel strength $\delta^*$. An approximate expression for the critical fuel strength is given as (see Appendix~\ref{sec:analytics} for the derivation)
\begin{equation}
    \delta^* \simeq k_BT \log{\left( 1 + \frac{k^\RN{1}_1}{k^\RN{2}_1}\Theta \right) } \, , \label{eq:ana_deltastar}
\end{equation}
where $\Theta$ depends solely on thermodynamic parameters (see Eq.~\eqref{eq:crit_interactions}). We see that the bifurcation is favored when the reactions are faster in the dense phase compared to the dilute phase ($k^\RN{2}_1 > k^\RN{1}_1$).
Moreover, the prerequisite for the bifurcation ($\Theta>0$) sets a restriction to certain phase diagram types.
For example, the bifurcation can occur when the 
 product $B$ of the fueled reaction  is
less soluble than its precursor $A$, for which $\Theta>0$.  
Furthermore, $\Theta$ never approaches zero implying that fuel is always necessary ($\delta>0$) for the emergence of bistability. 


We further find that bistability is only possible 
in a range of total volume fractions
$\bar{\psi}$-values and this range increases with fuel strength $\delta$ (orange domain in Fig.~\ref{fig:2}(d,e)).
The analytic expression for the critical fuel strength $\delta^*$ given in Eq.~\eqref{eq:ana_deltastar} agrees well with the complete numerical solution close to the active droplet domain in Fig.~\ref{fig:2}(e). 
As the total volume fraction $\bar{\psi}$ enters the binodal for a fixed fuel strength $\delta>\delta^*$, a jump occurs in both phase volume and average composition (Fig.~\ref{fig:2}(d)), unlike the case for $\delta < \delta^*$.

The onset of bistability can be understood through a normal form obtained from Taylor expanding the dynamic Eq.~\eqref{eq:dt_xi_bar} for the reaction extent $\bar{\xi}$. By expanding the phase-separated and homogeneous case separately, at the points $(\bar{\xi}_0,\bar{\psi}_0)$ and $(\bar{\xi}_2,\bar{\psi}_0)$ respectively, we find (derivation see Appendix~\ref{sec:effective_bistab}),
\begin{equation}
   \partial_t\bar{\xi} \simeq
\begin{cases}
-\left(\bar{\xi} - \bar{\xi}_0\right)h_1,\qquad\qquad\qquad\,\,\,\quad \,\bar{\xi} \leq \xi^\RN{1},\\
 \bar{r}_1(\bar{\xi}_2, \bar{\psi}_0) - \left(\bar{\xi} - \bar{\xi}_2\right)^2\,h_2\quad\qquad \bar{\xi} > \xi^\RN{1}\,,\label{eq:effective_equations}
\end{cases}
\end{equation}
with the coefficients $h_1$ and $h_2$ given in Eq.~\eqref{eq:h_1} and \eqref{eq:h_2} respectively, $\bar{r}_1$ is the phase-averaged reaction rate evaluated at the phase-separated expansion point, and $\xi^\RN{1}$ is the reaction extent at the dilute branch of the binodal for the given $\bar{\psi}_0$. As seen in Fig.~\ref{fig:2}(b), for homogeneous values of the reaction extent ($\bar{\xi}\leq\xi^\RN{1}$), the reaction rate is linear and stable ($h_1>0$) around its steady point $\bar{\xi}_0$. For phase-separated values of the reaction extent ($\bar{\xi}>\xi^\RN{1}$), a second-order expansion is necessary to capture the behaviour for strong fueling $\delta$. 
Since the normal form in the phase-separated domain corresponds to a concave parabola ($h_2>0$), the requirement for bistability is a positive offset value  at the expansion point: 
\begin{equation}
    \bar{r}_1 = r^\rightharpoonup_1\left(\overline{\exp{\frac{\delta}{k_BT}} - \exp{-\frac{\Delta \mu}{k_BT}}}\right) >0 
    \, .
    \label{eq:r_bar_rewrite}
\end{equation}
This equation shows that the pitchfork bifurcation occurs when the thermodynamic force $(\Delta \mu)$ is dominated by  the non-equilibrium driving $(\delta)$. 
The bar denotes a phase average (Eq.~\eqref{eq:phase_avg}).
We note that for $\bar{\psi}_0$ values with a homogeneous steady point, the chemical potential difference is negative when phase separated ($\Delta\mu<0$).


\begin{figure}[tb]
    \centering
      \makebox[\textwidth][c]{
      \includegraphics[width=\textwidth]{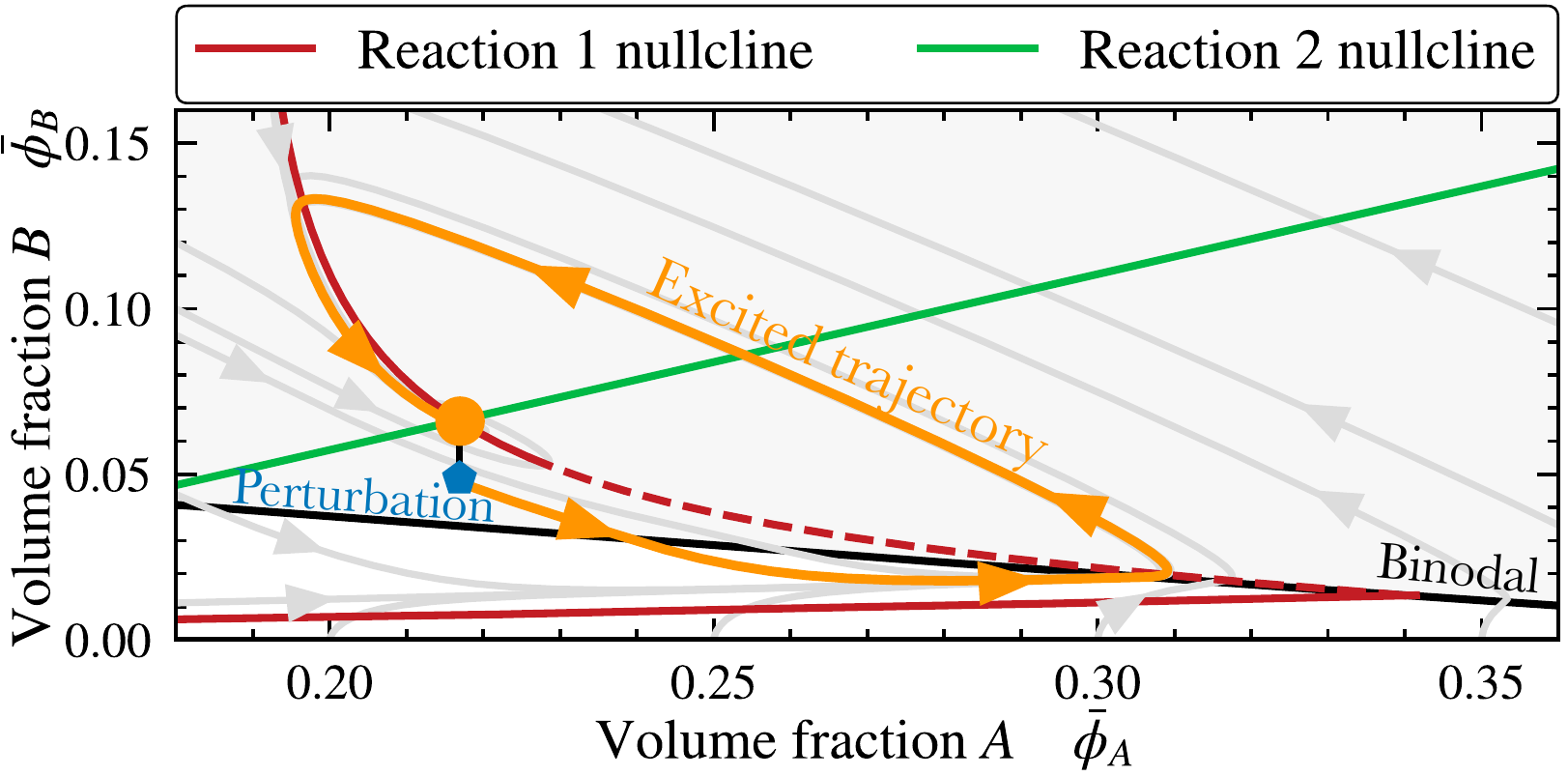}}
     \caption{
     \textbf{Excitable active droplet:}
     When considering two chemical reactions, reaction $r_1$ (Eq.~\eqref{eq:r_1_pathway}) and reaction $r_2$  (Eq.~\eqref{eq:r_2_pathway}), the active droplet system becomes excitable.    
     A compositional perturbation (blue pentagon)  can excite the system. As a result, the system follows an extended trajectory (orange)  with  changes in average composition far beyond the initial  perturbation. 
     During this excited trajectory, the system has a refractory period when it is no longer excitable. It relaxes back to the NESS, where the reaction nullclines for both reactions intersect (green and red lines).   
     The active droplet system can  be considered as bistable excitable media.
     }
     \label{fig:3}
\end{figure}

\begin{figure*}[tb]
    \centering
    \makebox[\textwidth][c]{
    \includegraphics[width=\textwidth]{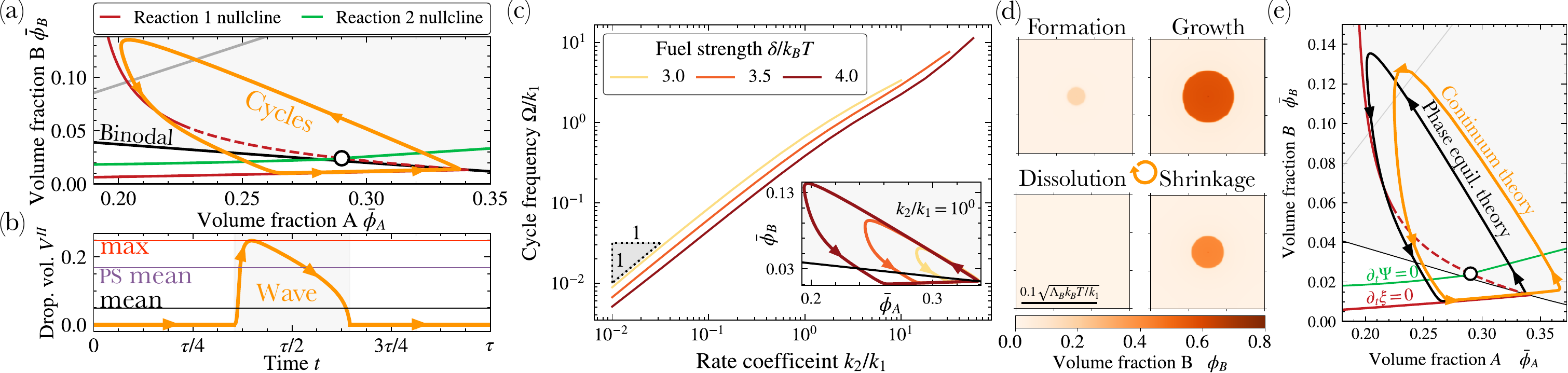}}
     \caption{
     \textbf{Self-sustained cycles of droplet formation and dissolution:} 
     (a) 
    For cases where the intersection between the two reaction nullclines (green and red lines) occurs for unstable droplet volumes (white dot), it cannot concomitantly satisfy its two steady-state conditions.
     The result is self-sustained oscillations of an active droplet cycling between formation and dissolution.
      (b) The droplet continuously forms and dissolves in time. When nucleated at the same position,  
      droplet formation and dissolution correspond to  a standing wave. 
      (c) The cycle frequency $\Omega$ is weakly affected by the fuel strength $\delta$. However, 
      the cyclic trajectories in phase space (inset) are significantly bigger, indicating that the compositional changes are much more pronounced. 
      The cycle frequency follows a power-law with the reaction rate coefficient $k_2$, $\Omega \propto k_2$, due to the fuel making $r_1$ relax much faster than $r_2$ upon formation and dissolution. 
      (d, e) We confirm active droplets' formation and dissolution cycles by solving the continuum equation for the volume fraction fields (Eq.~\eqref{eq:phi_dot}). 
      We find a good agreement of the phase space trajectory of the spatially averaged volume fractions for phase equilibrium theory (Eq.~\eqref{eq:kinetic_evolve}), where the main discrepancy originates from the droplet's nucleation time. An animation corresponding to panel (d) is included in the Supplementary Information.}
     \label{fig:4}
\end{figure*}

\textit{Excitable active droplet:} Excitable systems can lead to wave propagation upon a perturbation, followed by  a refractory period in which the system is no longer~\cite{cross1993pattern, turingChemicalBasisMorphogenesis1997,epsteinNonlinearChemicalDynamics1996, cross2009pattern}. Here, we show that the bistable system discussed in the previous paragraph becomes excitable upon adding a second reaction $r_2$  that allows the otherwise conserved quantity $\bar{\psi}$ to change in time,
\begin{equation}
r_2: \quad \text{A} + \text{B} \ce{<=>[$k_2$]} 2\text{C} \,. \label{eq:r_2} 
\end{equation}
The reaction obeys detailed balance of the rates
\begin{equation}
\label{eq:r_2_pathway}
    r_2=k_2\left( \exp{\frac{2\mu_C}{k_BT}} - \exp{\frac{\mu_A+\mu_B}{k_BT}}\right)\, .
\end{equation}
Thus, the net reaction rates $r_i$ in Eq.~\eqref{eq:phi_dot} and Eqs.~\eqref{eq:kinetic_evolve} is $r_A=r_2-r_1$ and $r_B=r_2+r_1$. 
The dynamics of the average reaction extent and the total volume fraction (Eq.~\eqref{eq:tot_vol_frac}) becomes
\begin{equation}
\label{eq:barxi_barpsi_1_2}
\partial_t\bar{\xi}=\bar{r}_1\, ,\qquad\quad\partial_t\bar{\psi}=\bar{r}_2\,.
\end{equation}

For a bistable droplet ($\delta>\delta^*$, Eq.~\eqref{eq:ana_deltastar}), we find that a compositional perturbation which destabilizes the droplet, resulting in an excited trajectory in the phase diagram (Fig.~\ref{fig:3}).
Such a perturbation leads to the dissolution of the droplet. Once dissolved, the reactions drive the average composition into the binodal, nucleating an active droplet, and returning the system to its initial non-equilibrium steady state. Along its excited trajectory, the system exhibits a refractory period when it is no longer excitable. The resulting evolution of the droplet volume in time is reminiscent of a wave pulse. Alternatively, bistable systems at a homogeneous steady state can be excited upon a perturbation that nucleates an active droplet; see Appendix~\ref{sec:excited_homo} for further discussions.

\textit{Oscillating active droplet:}
When the excitable fixed point becomes unstable, the excitable system undergoes cycles of formation and dissolution of the active droplet. 
Graphically speaking, destabilization corresponds to  
both nullclines of $r_1$ and $r_2$ intersect inside the phase-separated domain of the phase diagram.
As a result, a tiny droplet grows because the phase-dependent fueling  favors the turnover from $A$ to $B$,  increasing its volume (Fig.~\ref{fig:4}(b)). These compositional changes make the second net reaction rate $\bar{r}_2$ flip sign, decreasing the total volume fraction $\bar{\psi}$. This reduction in $\bar{\psi}$ lowers the droplet volume until crossing the unstable branch of the pitchfork bifurcation (Fig.~\ref{fig:2}(c,d), dashed lines). 
Consequently, the droplet dissolves, and the effect of fueling approximately disappears.
The lack of fuel reverts the direction of the net reaction rate $\bar{r}_1$, converting now $B$ back to $A$, leading to a flipped sign of $\bar{r}_2$. 
A droplet nucleates, reaching its initial state. The cycle of formation and dissolution of the active droplet continues indefinitely.

The transitions between an excitable system (Fig.~\ref{fig:3} and a system with oscillations of an active droplet (Fig.~\ref{fig:4}(a)) occur through a Hopf bifurcation~\cite{cross1993pattern, marsdenHopfBifurcationIts2012,caraballoStochasticpitchforkBifurcation2001, cross2009pattern}. At the Hopf bifurcation, eigenvalues of the linearization around the fixed point become imaginary. 
The transition to oscillations can, for example, be triggered by changing the reference chemical potential of component $C$, as defined in Eq.~\eqref{eq:decomp_chempot}, or by extending the unstable domain of $r_1$ by increasing the fuel strength $\delta$.

The Hopf bifurcation can be understood through the system's
normal form (Eqs.~\eqref{eq:barxi_barpsi_1_2}).
It is obtained by performing a Taylor expansion on both sides of the binodal (see Appendix~\ref{sec:effective_active} for the derivation), yielding
\begin{subequations}  \label{eq:effective_active} 
\begin{align}
\partial_t\bar{\psi} &\simeq -\left(\bar{\xi}-\bar{\xi}_1\right)f_\xi - \left(\bar{\psi} - \bar{\psi}_1\right)f_\psi\,, \\
\partial_t\bar{\xi} &\simeq 
\begin{cases}
   -\left(\bar{\psi} - \bar{\psi}_0\right)h_\psi -  \left(\bar{\xi} - \bar{\xi}_0\right)h_\xi\, ,\,\,\,\,\qquad \bar{\xi} \leq \xi^\RN{1},\\
   \bar{r}_1 - \left(\bar{\psi} - \bar{\psi}_1\right)\tilde{h}_\psi  - h_{\xi\xi}\left(\bar{\xi} - \bar{\xi}_2\right)^2\, , \, \, \, \bar{\xi} > \xi^\RN{1}\,.
\end{cases} 
\end{align}
\end{subequations} 
The key difference to the normal form with a single reaction $r_1$ (Eq.~\eqref{eq:effective_equations}) is that $\bar{\psi}$ is no longer a conserved quantity and changes during the kinetics. The consequences are coupling terms between the two dynamic quantities $\bar{\psi}(t)$ and $\bar{\xi}(t)$, characterized by   the coefficients $f_{\xi}$, $h_{\psi}$ and $\tilde{h}_{\psi}$ (definitions see Eqs.~\eqref{eq:def_rhos}, \eqref{eq:h_psi}, and \eqref{eq:h_psi_tilde}).
We find that $f_{\xi}>0$, $h_{\psi}<0$, and $\tilde{h}_{\psi}>0$,  leads 
to negative feedback  
between $\bar{\psi}(t)$ and $\bar{\xi}(t)$ 
(see discussion in Appendix~\ref{sec:effective_active}).
The signs of $f_{\xi}$ and $h_{\psi}$ are set by thermodynamics, while the difference in sign between $\tilde{h}_{\psi}$ and $h_{\psi}$ is ensured from the fuel. Negative feedback loops, together with the instability at the intersection of the two nullclines, represent a well-established fundamental ingredient for the emergence of oscillations in non-linear systems~\cite{cross1993pattern,strogatz2018nonlinear}. For cases where the two quantities can relax to a stable fixed point, i.e., $\bar{\psi}_0=\bar{\psi}_1$ and $\bar{\xi}_0=\bar{\xi}_1$ (homogeneous), or alternatively $\bar{\xi}_2=\bar{\xi}_1$ (phase-separated), the system does not oscillate, however can remain excitable.

The oscillation frequency 
is dominantly determined by the reaction rate coefficient $k_2^{}$ of the second reaction~\eqref{eq:r_2_pathway}.
This property is apparent in the scaling of the
cycle frequency $\Omega\propto k_2$, as shown in Fig.~\ref{fig:4}(c). 
The dominant role of $k_2$ can be explained as follows:
The first reaction~\eqref{eq:r_1_pathway} is fast since phase separation switches on the fueled forward pathway converting $A$ to $B$, a prerequisite for having an active droplet. To complete the cycle, the slower unfueled reaction 2 with $k_2\ll k_1 \exp\{\delta/(k_BT)\}$ becomes rate limiting, thereby  determining the cycle frequency $\Omega$. 
This separation of time-scales is consistent with 
the fuel strength $\delta$ 
having only little effects on the cycle frequency $\Omega$
(different colors in Fig.~\ref{fig:4}(c)).
Notably, the compositional changes along a cycle are more pronounced for larger fuel strength $\delta$ (Fig.~\ref{fig:4}(c), inset).

Consistently, we find formation-dissolution cycles when numerically solving the  continuous dynamic equations for the volume fraction fields $\phi_i(\bm{x},t)$ ($i=A,B$) (Eq.~\eqref{eq:phi_dot}).
Fig.~\ref{fig:4}(d) depicts representative snapshots of the volume fraction fields indicating the formation, growth, shrinkage, and dissolution of an active droplet over the period $(2\pi/\Omega)$.
Fig.~\ref{fig:4}(e) shows the corresponding phase space trajectory, which agrees well with the fast-diffusion limit. The self-sustaining oscillations are, therefore, well captured by the dynamic equations derived in the fast-diffusion limit which paved the way to determine the normal forms of the pitchfork bifurcation (Eq.~\eqref{eq:effective_equations}) of an active droplet and the Hopf bifurcation (Eq.~\eqref{eq:effective_active}) to cycles of formation and dissolution of an active droplet.

\textit{Entropy production of an active droplet:} The entropy production characterizes the system's dissipation and serves as a measure of how far a system is away from thermodynamic equilibrium~\cite{julicherModelingMolecularMotors1997b}. For an active droplet, the entropy production rate $\dot{S}$ at the non-equilibrium steady state (derivation see Appendix~\ref{sec:fuel}) scales
\begin{equation}
{\dot{S}_\text{NESS}} \propto 
{k_B V} k_1  
\left( \exp{\frac{\mu_B}{k_BT}} - \exp{\frac{\mu_A}{k_BT}}\right)\, .    \label{eq:entropy_prod_NESS}
\end{equation}

\begin{figure}[tb]
    \centering
      \makebox[\textwidth][c]{
      \includegraphics[width=\textwidth]{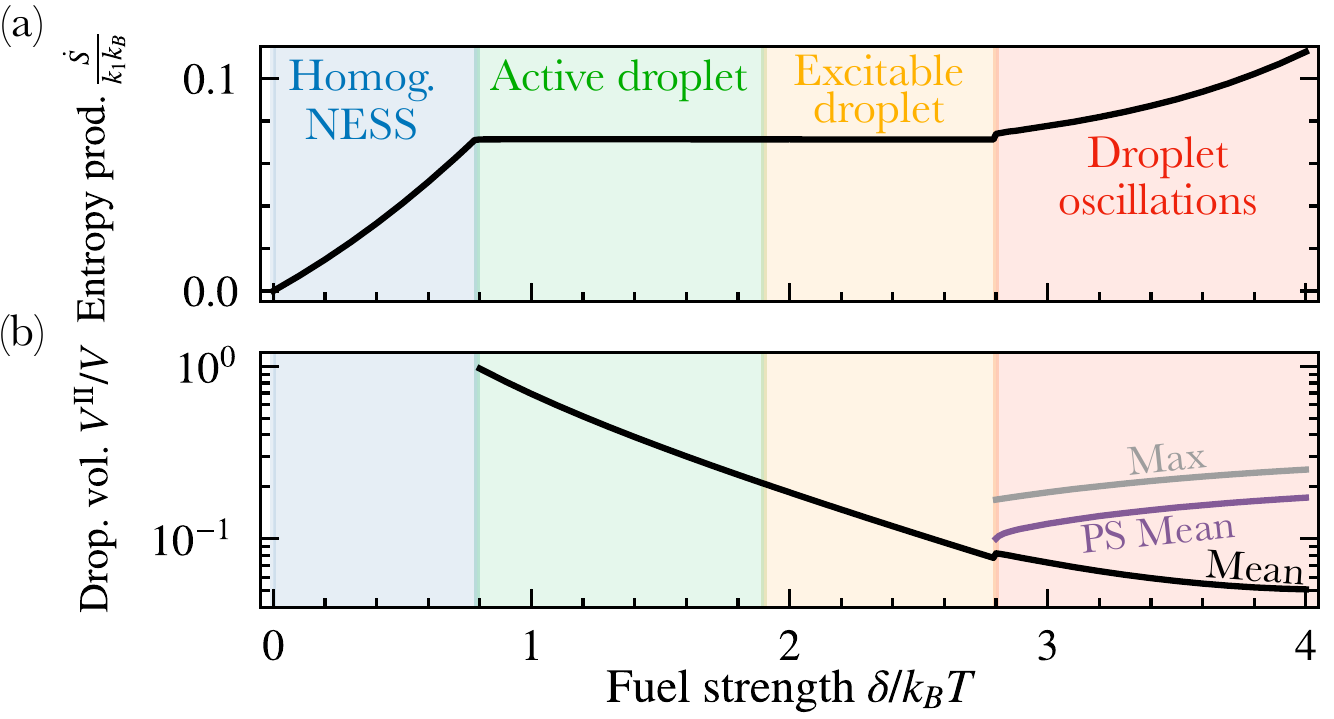}}
     \caption{
     \textbf{
     Discontinuous entropy production at the onset of oscillations.}
     Upon increasing the fuel strength $\delta$, homogeneous NESS transits to the active droplet state with a stationary droplet volume with a kink in the entropy production $\dot S$. 
    Increasing $\delta$ further decreases the droplet volume (b) but keeps the entropy production (a) constant; further discussion see Appendix~\ref{sec:fuel}. At the onset of formation-dissolution cycles, the entropy production $\dot S$ jumps (transition between the green and yellow domain in (a)). The volume  averaged over a period jumps slightly, while the peak of the wave (Max) and its mean volume when phase separated (PS Mean) (see Fig.~\ref{fig:4}(b)) show a significant jump, suggesting both as ideal read-outs in experiments of self-sustained cycles. The analytic prediction for the critical fuel strength is the horizontal orange line. 
     }
     \label{fig:5}
\end{figure}

As the chemical reaction $r_2$ (Eq.~\eqref{eq:r_2})
obeys detailed balance of the rates,
the chemical potentials $\mu_A$ and $\mu_B$ are fixed by the tie-line selected by the nullcline of $r_2$  (green line in Fig.~\ref{fig:3} and Fig.~\ref{fig:4}(a)). Thus, the entropy production for active droplets is independent of the fuel strength $\delta$, and is instead a constant set by the properties of the non-dissipative chemical reaction 2. For a fixed $\Delta\mu=\mu_A-\mu_B$, we see from Eq.~\eqref{eq:NESS} that an increase in fuel strength $\delta$ decreases the droplet volume. This reduced volume exactly balances the increased dissipation from a stronger fuel, keeping the entropy production constant, as seen in Fig.~\ref{fig:5}. We note again that without a second reaction, which obeys detailed balance, the entropy productions will increase with the fuel strength $\delta$~\cite{bauermannEnergyMatterSupply2022d}.

Increasing the fuel strength further, the droplet transitions from an active droplet to an excitable droplet, where the steady state is stable. Upon further increasing the fuel strength $\delta$, cycles of formation and dissolution emerge. The transition from stationary, excitable droplets to oscillations creates a discontinuous jump in the entropy production rate $\dot{S}$. Over a cycle, the average entropy production rate $ \langle \dot{S} \rangle\equiv ({2\pi}/{\Omega})\int_0^{2\pi/\Omega}\dot{S}\, \text{d}t$, continues to grow with fuel strength $\delta$. As the entropy production rate jumps, so does the average droplet volume; see Fig.~\ref{fig:5}(b). An even larger jump occurs when comparing the average droplet volume when phase-separated (PS mean) and the maximum droplet volume over a cycle (Max). Formation-dissolution cycles thus permit a discontinuous increase in entropy production and larger droplet volumes as the system is driven away from equilibrium.

\textit{Discussion:} 
Cyclic reaction networks in homogeneous systems, such as the Brusselator~\cite{prigogineSymmetryBreakingInstabilitiesDissipative1967,prigogineSymmetryBreakingInstabilities1968}, rely on a complex non-linear reaction network of many components with a large variety of time scales~\cite{novakDesignPrinciplesBiochemical2008,semenovRationalDesignFunctional2015}. The oscillations studied here occur with only two independent components. 
Our phase-separated system can instead use the phase boundary to create a simple cyclic network by effectively increasing the number of degrees of freedom through the phase-dependent concentrations and a cyclic network with a single detailed-balance breaking reaction,
$A^\RN{1} \textcolor{black}{\rightarrow} A^\RN{2}\textcolor{black}{\rightarrow} B^\RN{2}\textcolor{black}{\rightarrow} B^\RN{1}\textcolor{black}{\rightarrow} A^\RN{1}$, where diffusive exchange between the phases I and II act as effective reaction steps.  

Bistability in homogeneous chemical systems relies on higher-order reaction networks to achieve a cubic normal form~\cite{epsteinNonlinearChemicalDynamics1996,novakDesignPrinciplesBiochemical2008}. Our phase-separated system gives rise to
the effective properties of a cubic form through
two phases with a piece-wise combination of a linear and quadratic function (Eq.~\eqref{eq:effective_equations}).
In other words, phase separation enables three stable points despite only two independent chemical states.
The bistability found in Fig.~\ref{fig:2} is, therefore, not a consequence of the complexity of the reaction network as such, it is a property of active droplets. Specifically, bistability arises from the coupling between phase separation and detailed-balance-breaking chemical reactions far from equilibrium, which we have shown can give rise to excitability and self-sustaining chemical oscillations. 

The relevance of thermodynamic interactions among components for the bistability enables the same chemical reaction network to give fundamentally different oscillations. Changing thermodynamic parameters leads to different binodals, strongly altering the composition changes during a cycle. We find, for example, self-sustained oscillations of the active droplet's volumes; see Appendix~\ref{sec:volume_oscillations}.

\textit{Conclusion:} In our work, we present a minimal theoretical model to achieve bistability, excitability, and oscillations in phase-separated systems with chemical reaction maintained away from equilibrium.  We show that this non-equilibrium driving enables a pitchfork bifurcation in droplet volumes with two locally stable chemical states (bistability). By including an additional reaction, the system becomes excitable. For cases where the non-equilibrium steady state is on either of the stable branches, perturbations trigger an excited trajectory in composition, during which the system exhibits a refractory period. Upon enhancing the non-equilibrium driving, a Hopf bifurcation occurs to a self-sustained oscillating droplet undergoing indefinite formation and dissolution cycles.

Excitable and oscillating active droplets can be realized experimentally in multiple ways. The general requirement is a phase-separating system with a phase-dependent fueled reaction pathway 
and a second chemical reaction that alters the quantity conserved by the fueled reaction. 
Moreover, the product of the fueled pathway should phase-separate more strongly than the precursor reactant. 
A potential experimental system could rely on  transient post-translational modifications of proteins, where phase separation is activated through arginine methylation or phosphorylation~\cite{luo2021regulation}.
Alternatively, various synthetic systems were recently proposed that use chemical fuels to transiently activate thermodynamically unstable products for active droplet formation~\cite{donau2023chemistry}, 
either through  phosphorylation of peptides~\cite{poprawa2024active} or 
fuel-driven condensing reactions~\cite{bergmannLiquidSphericalShells2023d,sastre2025size}.

Our work is relevant for non-equilibrium systems that are phase-separated  and composed of evolving  oligomer sequences~\cite{bartolucciInterplayBiomolecularAssembly2024c,haugerudTheorySequenceSelection2024}. Such systems were proposed as simple prebiotic compartments at the Origin of Life~\cite{ianeselli2023physical}. Cycles of formation and dissolution can mediate a search mechanism in sequence space, reminiscent of stochastic resetting~\cite{evansDiffusionStochasticResetting2011} of the dense environment in which selection can occur~\cite{haugerudTheorySequenceSelection2024}. Resetting is beneficial in search processes of specific sequences in a high-dimensional sequence space~\cite{evansStochasticResettingApplications2020}, allowing their self-replication, for example. 


The emergence of oscillations through non-equilibrium phase separation makes them an ideal functional module for \textit{de novo} life systems~\cite{kriebischRoadmapSynthesisLife2025}. Robust self-sustained chemical oscillations are crucial for \textit{de novo} life to facilitate clock-like functions similiar to extant life, such as the circadian rhythm and cell cycle~\cite{zehringPelementTransformationPeriod1984,bargielloRestorationCircadianBehavioural1984,dunlapMolecularBasesCircadian1999,goldbeterBiochemicalOscillationsCellular1996}, and could thereby provide in other biological functions such as DNA repair~\cite{hsiehPlausibleRobustBiological2024,heltbergEnhancedDNARepair2022,heltbergCoupledOscillatorCooperativity2023,kriebischRoadmapSynthesisLife2025}. 

\textit{Acknowledgements:}
We thank Leonardo Silva Dias for careful feedback on the manuscript and his recommendations for references on chemical oscillators. 
C.\ A.\ Weber acknowledges the European Research Council (ERC) under the European Union's Horizon 2020 research and innovation program (Fuelled Life,  Grant Number 949021), and the SPP 2191 “Molecular Mechanisms of Functional Phase Separation” of the German Science Foundation for financial support.

\bibliography{cycles.bib}

\appendix
\section{Free energy and chemical potentials}\label{seq:free_energy_and_chempot}
The chemical potentials that govern the dynamics of the system (Eq.~\eqref{eq:phi_dot}) are calculated by changes in the mean-field Gibbs free energy $G$ per volume $V$, where we use the Flory-Huggins free energy density \cite{bauermannChemicalKineticsMass2022d,adame-aranaLiquidPhaseSeparation2020a},
\begin{align}
    \frac{G(T, p, \{\phi_k\})}{V} = k_\text{B}T &\sum_{i=0}^M \left[\frac{\phi_i}{\nu_i}\log{\phi_i} + \frac{\omega_i\phi_i}{\nu_ik_BT}\right] \nonumber\\ + \frac{1}{2\nu_0} &\sum_{i=0}^M\sum_{j=0}^M \chi_{ij}\phi_i\phi_j+ p \, .\label{eq:gibbs}
\end{align}
Here, $\omega_i$ denotes the internal energies, $\chi_{ij}$ characterize the interaction strength between component $i$ and $j$, and $p$ is the pressure. The volume fractions are defined as $\phi_i\equiv\nu_iN_i/V$, where $\nu_i$ is the molecular volume of component $i$, and $N_i$ is the number of particles of that type. From the change in Gibbs free energy with respect to particle number $N_i$, we find the chemical potential ${\mu}_{i} = \partial{G}/{\partial N_{i}}\vert_{N_{j\neq i}}$ to be on the form 
\begin{equation}
    \frac{\mu_i}{\nu_i} = \left(\pdv{g}{\phi_i}\right)_{T,p,\phi_{j\neq i}} + \left(g - \sum_{j}\phi_j\left(\pdv{g}{\phi_j}\right)_{T,p,\phi_{k\neq j}} \right)\,, \label{eq:rewrite_mu}
\end{equation}
where $g$ is the Gibbs free energy per volume $G/V$ given in Eq.~\eqref{eq:gibbs}. The first term in Eq.~\eqref{eq:rewrite_mu} represents the change in free energy when adding a particle due to compositional changes, while the second term represents the change due to the volume change of the system. Evaluating the derivatives grants us an expression for the chemical potentials
\begin{equation}
    \mu_i = k_BT\left(\log{\phi_i}+1\right)+\omega_i + \nu_i\left(p-\Gamma\right) + \frac{\nu_i}{\nu_0}\sum_{j}\chi_{ij}\phi_j, \label{eq:chem_pot}
\end{equation}
where we have defined 
\begin{equation}
    \Gamma = \frac{1}{2\nu_0}\sum_{ij}\chi_{ij}\phi_i\phi_j + k_BT\sum_i\frac{\phi_i}{\nu_i}.
\end{equation}
The chemical potential can be separated into a composition-independent contribution $\mu_i^0$ known as the reference chemical potential, and a composition-dependent activity coefficients $\gamma_i$, such that
\begin{equation}
    \mu_i = \mu_i^0(T,p) + k_BT\,\log{\left(\phi_i\gamma_i(\{\phi_j\}, T, p) \right)}\,. \label{eq:decomp_chempot}
\end{equation}
The activity coefficient approaches a constant in the limit of weak interactions or solvent-rich systems, leading to chemical reaction rates agreeing with the law of mass-action \cite{bauermannChemicalKineticsMass2022d}. When solving the continuum equation, a mobility of $\Lambda_i = \Lambda_i^0\phi_i$ is chosen to retrieve the diffusion equation in the aforementioned limit. 

For the spatial profile in Fig.~\ref{fig:4}(d) found from solving the full continuum equation (Eq.~\eqref{eq:phi_dot}) the free energy $G$ is calculated from
\begin{align}
\begin{split}
    G = \int_V \text{d}^3\bm{x}\, &\Bigg[ g(\phi_i(\bm{x})) + \frac{\kappa_A}{2}\left(\nabla\phi_A\right)^2 \\ &+ \frac{\kappa_B}{2}\left(\nabla\phi_B\right)^2   + U(\bm{x})(\phi_A + \phi_B)\Bigg]\,. \label{eq:gibbs_spatial}
\end{split}
\end{align}
Here, $g$ is the Gibbs-free energy density, i.e., the right-hand side of Eq.~\eqref{eq:gibbs}, $\kappa_i$ is the free energy cost associated to gradients of component $i$, and $U$ is an external potential affecting $A$ and $B$, which is included to act as a nucleation potential. The chemical potential of component $i$ for this free energy is calculated from the functional derivative of $G$ with respect to volume fraction,
\begin{equation}
    \tilde{\mu}_i = \nu_i\frac{\delta G}{\delta \phi_i}\, .
\end{equation}
The inclusion of the free energy cost characterized by  $\kappa_i$ to the free energy yields an additional term in the chemical potential with respect to: Eq.~\eqref{eq:chem_pot}
\begin{equation}
    \tilde{\mu}_i = \mu_i - \kappa_i\nabla^2\phi_i + U\left[1-\delta_{i,C}\right] \, . \label{eq:chem_pot_spatial}
\end{equation}

By using the free energy in Eq.~\eqref{eq:gibbs}, 
there can be phase separation that gives rise to domains of very different values of the total volume fraction $\psi(\bm{x},t)=\phi_A(\bm{x},t)+\phi_B(\bm{x},t)$. 
The free energy cost parameter $\kappa_i$ is related to surface tension that leads to a droplet size-dependent Laplace pressure, increasing the respective volume fraction in each domain. 
Since the binodal displayed in Fig.~\ref{fig:4}(e) is calculated without such surface tension effects, it does not exactly correspond to the binodal for the continuum equation. Still, it serves as a good approximation when the droplet's radius is much larger than the capillary length~\cite{weberPhysicsActiveEmulsions2019d}.

The potential $U$ in Eq.~\eqref{eq:gibbs_spatial} and \eqref{eq:chem_pot_spatial} is included to create a nucleation site, which allows droplet nucleation in the meta-stable region inside the binodal without noise. This is done as we are not interested in probing the nucleation or nucleation-time scales in this investigation. The role of stochasticity will be discussed in future work. \par

\section{Reaction scheme with an explicit fuel component}\label{sec:fuel}
Detailed balance is broken by a fuel $\tilde{\delta}$ according to Eq.~\eqref{eq:r_2_pathway}. Breaking detailed balance in this fashion is equivalent to the combined effect of the two reactions:
\begin{equation}
    A\xrightleftharpoons{k_u}{} B\, ,\qquad A+F\xrightharpoonup{k_\text{f}}{} B\, ,
\end{equation}
where the former reaction obeys detailed balance of the rates while the latter breaks it. The reaction rates of the unfueled $(r_\text{u})$ and fueled reaction $(r_\text{f})$ are
\begin{align}
    r_\text{u} &= k_u\left[\exp{\frac{\mu_A}{k_BT}} - \exp{\frac{\mu_B}{k_BT}} \right], \label{eq:r_u}\\
    r_\text{f} &= k_\text{f}\exp{\frac{\mu_A+\mu_F}{k_BT}}\, , \label{eq:r_f}
\end{align}
such that $r_B = -r_A = r_\text{u}+r_\text{f}$. Performing the mapping between these rates and the coarse-grained rate studied (Eq. \eqref{eq:r}), we find 
\begin{equation}
    \tilde{\delta} = \log{\left(1+\frac{k_\text{f}}{k_u}\exp{\frac{\mu_F}{k_BT}}\right)}, \label{eq:delta_muF}
\end{equation}
where $\mu_F$ is the chemical potential of the fuel. Solving for the reaction rate coefficient of the fueled reaction $k_\text{f}$, we find
\begin{equation}
    k_\text{f} = k_u\exp{-\frac{\mu_F}{k_BT}}\left(\exp{\frac{\tilde{\delta}}{k_BT}}-1\right), \label{eq:k_f}
\end{equation}
with $k_u = k_1$.
Thus, the fuel's chemical potential can be regarded as phase-independent, meaning it satisfies the conditions of phase coexistence, and the phase-dependent fuel strength originates from the composition dependence of the rate coefficient $k_\text{f}$. For the studied case of $\tilde{\delta}^{\RN{1}}=0$ and $\tilde{\delta}^{\RN{2}}=\delta$, this is equivalent to the fueled reaction having a reaction rate of zero in the dilute phase, and a non-zero value according to Eq.~\eqref{eq:k_f} in the dense phase.

The detailed-balance breaking reaction produces entropy. The entropy production originates from the fact that the fuel is chemostatted, implying an entropy-producing boundary flux to maintain a chemostatted fuel. The entropy production rate reads \cite{julicherGenericTheoryColloidal2009a,bauermannEnergyMatterSupply2022d}
\begin{equation}
    -T\partial_t s = \partial_t g + j_F\,,
\end{equation}
The flux of fuel into the system $j_F$ balances the loss of fuel by the fueled reaction, $j_F = - \mu_Fr_F/\nu_F$, where we now explicitly keep track of the chemostatted fuel with a constant chemical potential $\mu_F$ and molecular volume $\nu_F$. In a non-equilibrium steady state $\partial_t g=0$, and integrates to zero over a closed cycle, such that all entropy production originates from the dissipative fueled reaction $r_F$. From the fueled reaction rate (Eq.~\eqref{eq:r_f}), we find 
\begin{equation}
    \frac{\dot{s}}{k_B} = \frac{\mu_Fk_\text{f}}{k_BT\nu_F}\exp{\frac{\mu_A+\mu_F}{k_BT}} \, .    \label{eq:entropy_prod_rate_homogeneous}
\end{equation}
Performing a volume integral to find the entropy production rate $\dot{S}=\int \dot s\,\text{d}^3x$ yields
\begin{equation}
    \frac{\dot{S}}{k_B} = \frac{\mu_F}{k_BT\nu_F}\exp{\frac{\mu_A+\mu_F}{k_BT}}\left(V^\RN{1}k_\text{f}^\RN{1} + V^\RN{2}k_\text{f}^\RN{2}\right) .    \label{eq:entropy_prod}
\end{equation}
The entropy production over a cycle is calculated from integrating this expression over a period
\begin{equation}
\langle \dot{S} \rangle=\frac{\Omega}{2\pi}\int_0^{2\pi/\Omega}\dot{S}\,\text{d}t\,.
\end{equation}
At a non-equilibrium steady state, we can use the expression of the droplet volume $V^\RN{2}$ (Eq.~\eqref{eq:NESS}) and the relation between fuel strength $\tilde{\delta}$ and $k_\text{f}$ in Eq.~\eqref{eq:k_f}, where $\tilde{\delta}^\RN{1}=0$ and $\tilde{\delta}^\RN{2}=\delta$. The entropy production at the non-equilibrium steady state simplifies to
\begin{equation}
    \frac{\dot{S}}{k_B} = \frac{\mu_Fk_1V}{k_BT\nu_F}\left( \exp{\frac{\mu_B}{k_BT}} - \exp{\frac{\mu_A}{k_BT}} \right)\, .
\end{equation}
Since the fuel is aiding the transition from $A$ to $B$ $(\mu_F>0)$, we have that $\mu_A<\mu_B$, ensuring that the entropy production is positive. For the case where the fuel strength goes to zero, the tie-line corresponding to chemical equilibrium, i.e., $\mu_A=\mu_B$, will be selected, making the entropy production rate zero. Interestingly, the entropy production is independent of the fuel strength and the phase volumes. 

\section{Kinetics at phase equilibrium}\label{seq:SJ_method}

In this section, we follow the work of Bauermann et al.~\cite{bauermannChemicalKineticsMass2022d} to derive the expression for the diffusive flux between the two phases, $j_i^{\RN{1}/\RN{2}}$, in Eqs.~\eqref{eq:kinetic_evolve}. This approach to phase-separated kinetics is valid when the system is at phase equilibrium throughout the kinetics, i.e., if the system size $L$ is much smaller than the reaction-diffusion length; $k_1 \ll \Lambda_ik_BT/L^2$. For phase-separated systems, this implies that the concentration profiles of each component in both phases are uniform $\phi_i^{\RN{1}/\RN{2}}$, and the interface is infinitely thin (the $\kappa$-terms in Eq.~\eqref{eq:gibbs_spatial} are ignored). 

For two phases to remain at phase coexistence, the chemical potentials of all $M-1$ components in the two phases change equally in time,
\begin{equation}
    \partial_t \mu_i^\RN{1} = \partial_t \mu_i^\RN{2}, \label{eq:equal_chem}
\end{equation}
and are initiated at equal values of the chemical potentials $\mu_i^\RN{1}(t=0) = \mu_i^\RN{2}(t=0)$, but different compositions. Using the chain rule, the time derivative of a chemical potential can be written as 
\begin{equation}
     \partial_t \mu_i^\alpha = \sum_{k=1}^M \dot{\phi_k^\alpha}\pdv{\mu_i^\alpha}{\phi_k} \, , \label{eq:mu_dot}
\end{equation}
where $\alpha$ is the phase index; $\alpha\in\{\RN{1}, \RN{2}\}$.
Incompressible systems, where the molecular volumes are independent of composition $\text{d}_t\nu_i=0$, follow Eqs.~\eqref{eq:kinetic_evolve}. Combining Eqs.~\eqref{eq:kinetic_evolve} with the expression for the volume change of the phases (Eq.~\eqref{eq:vol_change}) for volume conserving chemical reaction $(\sum_{i=1}^M r_i=0)$, we obtain
\begin{equation}
    \dot{\phi_k^\alpha} = r_k^{\alpha} - j_k^{\alpha} + \phi_k^{\alpha} \sum_{l=1}^M j_{l}^{\alpha} \, . \label{eq:kinetic_in_app}
\end{equation}
Such that the chemical potential change in Eq. \eqref{eq:mu_dot} takes the form
\begin{equation}
     \partial_t \mu_i^\alpha = \sum_{k=1}^M \left( r_k^{\alpha} - j_k^{\alpha}  + \phi_k^{\alpha} \sum_{l=1}^M j_l^{\alpha} \right)\pdv{\mu_i^\alpha}{\phi_k}\, ,
\end{equation}
which can be expressed in terms of the fluxes: 
\begin{align}
    \begin{split}
    \partial_t \mu_i^\alpha &= \sum_{k=1}^M j_k^\alpha\left(-\pdv{\mu_i^\alpha}{\phi_k} + \sum_{l=1}^M \phi_l^\alpha \pdv{\mu_i^\alpha}{\phi_l}\right)+  \sum_{k=1}^M \pdv{\mu_i^\alpha}{\phi_k}r_k^\alpha \, .\end{split}
\end{align}
The kinetic constraint on the chemical potentials (Eq.~\eqref{eq:equal_chem}) thus reads
\begin{align}
     &\sum_{k=1}^M j_k^\RN{1}\left\{ \sum_{l=1}^M \phi_l^\RN{1} \pdv{\mu_i^\RN{1}}{\phi_l^\RN{1}} -\pdv{\mu_i^\RN{1}}{\phi_k^\RN{1}} + \frac{V^\RN{1}}{V^\RN{2}}\left(\sum_{l=1}^M \phi_l^\RN{2} \pdv{\mu_i^\RN{2}}{\phi_l^\RN{2}} -\pdv{\mu_i^\RN{2}}{\phi_k^\RN{2}} \right) \right\} \nonumber \\ &= -\sum_{k=1}^M \pdv{\mu_i^\RN{1}}{\phi_k^\RN{1}}r_k^\RN{1} +  \sum_{k=1}^M \pdv{\mu_i^\RN{2}}{\phi_k^\RN{2}}r_k^\RN{2} \, ,\label{eq:chem_pot_constraint} 
\end{align}
where we have used the conservation of particle volume of the diffusive flux, 
\begin{equation}
    j_i^\RN{1}V^\RN{1} = -j_i^\RN{2}V^\RN{2} \, . \label{eq:particle_cons}
\end{equation}
This set of equations \eqref{eq:chem_pot_constraint} can be written as a matrix-vector equation
\begin{equation}
    A_{ik}j^\RN{1}_k = s_i,
\end{equation}
which can be solved for $j^\RN{1}_k$. The reaction rates in each phase act as source terms in $s_i$, while $A_{ik}$ specifies how the composition changes in each phase will affect the chemical potentials, constraining the diffusive fluxes between the phases to equalize the chemical potential change for all components in both phases.

\section{Parameters, numerical implimentation of continuum equation, and full phase diagrams}\label{seq:parameters_used}
For reproducibility, the parameters used to generate all figures are given in Table~\ref{tab:params}. We have assumed that the reaction rate coefficients are composition-independent $k_{1}^\RN{1}=k_{1}^\RN{2}=k_1$ and $k_{2}^\RN{1}=k_{2}^\RN{2}=k_2$. For the continuum simulation in Fig.~\ref{fig:4}(d,e), there are a few parameters in addition to the ones listed in Table~\ref{tab:params}, they are listed here instead: 
System size $L/\ell=0.15$, where $\ell$ is the reaction-diffusion length scale $\ell\equiv\sqrt{k_BT\Lambda/k_1}$, surface tensions $\kappa_A/\mathcal{K} = \kappa_B/\mathcal{K} 
 = 1.8\cdot10^{-5}$, where $\mathcal{K} = k_BT \ell^2$. The time-step $\delta t=10^{-7}/k_1$, and the number of grid points $N_x=N_y=250$. In order to retrieve the diffusion equation in the dilute limit $(\chi_{ij}=0)$, the mobilities scale linearly with the volume fractions according to $\Lambda_i = \Lambda_i^{(0)}\phi_i$. Lastly, the external potential is $U=-1.5k_BT\delta_{i,N_x/2}\delta_{j,N_y/2}$, where $i$ and $j$ are the spatial indices in the $x$ and $y$ direction respectively, and $\delta_{\alpha,\beta}$ is the Kronecker delta. The system was initialized with $\bar{\phi}_A = 0.32$ and $\bar{\phi}_A = 0.014$.

To solve the full spatial equation (Eq.~\eqref{eq:phi_dot}) needed to generate the data in Fig.~\ref{fig:4}(d) and (e), the spectral-method is employed~\cite{zhuCoarseningKineticsVariablemobility1999}. The spectral method solves the continuous dynamic Eqs.~\eqref{eq:phi_dot} in Fourier-space. Periodic boundary conditions were used.\par

\begin{table*}[]
 \caption{\textbf{Parameter choices for each figure:} Common for all figures in this work is that the molecular volumes are equal, $\nu_i=1$. All energy values in the table are given in units of $k_BT$ ($\chi_{ij}/k_BT$, $\omega_i/k_BT$ and $\delta/k_BT$). \label{tab:params}}
\begin{tabular}{@{}lllllllllllll@{}}
\toprule
Figure $\quad\qquad\quad$& $\chi_{AB}\quad$ & $\chi_{AC}\quad$ & $\chi_{BC}\qquad$ & $\omega_A\qquad$ & $\omega_B\qquad$  & $\omega_{C}\qquad$ & $k_2/k_1 \qquad$ & $\delta\qquad$ & $\kappa\qquad$ &  $\psi_{1/2}\qquad$ & $\bar{\psi}\qquad$ \\ \midrule
\ref{fig:2}(a, b, c) \& \ref{fig:appendix2}      &    0.0      &    2.01      &   3           &  -2.5      &  0.1   & -2.25 & --  & --- & $5\cdot10^{-4}$ & 0.25 & 0.16  \\
\ref{fig:2}(d)     &    0.0      &    2.01      &   3           &  -2.5      &  0.1   & -2.25 & --  & 4 & $5\cdot10^{-4}$ & 0.25 & --  \\
\ref{fig:2}(e)     &    0.0      &    2.01      &   3           &  -2.5      &  0.1   & -2.25 & --  & 4 & $5\cdot10^{-4}$ & 0.25 & --  \\
\ref{fig:3}     &    0.0      &    2.01      &   3           &  -1.5      &  0.1   & -1.68 & 5  & 4 & $5\cdot10^{-4}$ & 0.25 & --  \\
\ref{fig:4}(a,b,e) \& \ref{fig:appendix3}    &    0.0      &    2.01      &   3           &  -2.5      &  0.1   & -2.25 & 1.0  & 4 & $5\cdot10^{-4}$ & 0.25 & --\\
\ref{fig:4}(c)     &    0.0      &    2.01      &   3           &  -2.5      &  0.1   & -2.25 & --  & -- & $5\cdot10^{-4}$ & 0.25 & -- \\
\ref{fig:5}     &    0.0      &    2.01      &   3           &  -2.5      &  0.1   & -2.25 & 1.0  & --- & $5\cdot10^{-4}$ & 0.25 & --  \\
\ref{fig:3_appendix}     &    0.0      &    2.01      &   3           &  -2.5      &  0.1   & -1.925 & 5  & 4 & $5\cdot10^{-4}$ & 0.25 & --  \\
\ref{fig:appendix_intermediateDrops}     &    0.6      &    3.0      &   -0.6           &  -2.5      &  0.5   & -1.35 & 1.0  & 2.0 & 0.04 & 0.225 & --\\
\bottomrule
\end{tabular}
\end{table*}

\begin{figure}
    \centering
      \makebox[\textwidth][c]{
      \includegraphics[width=0.8\textwidth]{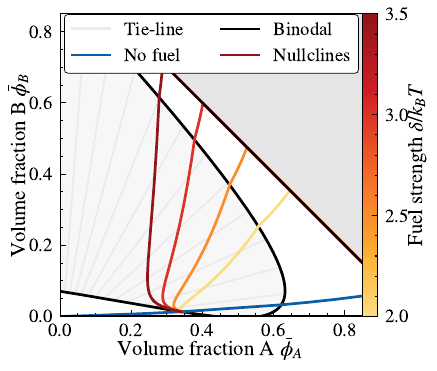}}
     \caption{
     \textbf{Full phase diagram:} The full phase space, instead of the subset displayed in Fig.~\ref{fig:2}(a), shows the entire binodal and how the unfueled system selects a single tie-line.}
     \label{fig:appendix2}
\end{figure}

\begin{figure}[tb]
    \centering
      \makebox[\textwidth][c]{
      \sidesubfloat[]{\includegraphics[width=0.7\textwidth]{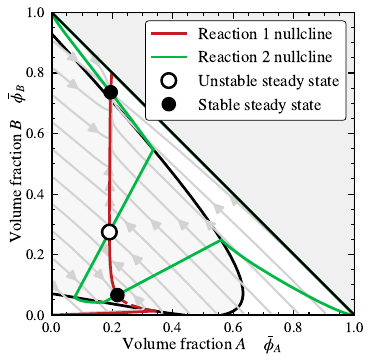}\label{fig:2a}}}\\
    \vspace{0.2cm}
      \makebox[\textwidth][c]{
      \sidesubfloat[]{\includegraphics[width=0.7\textwidth]{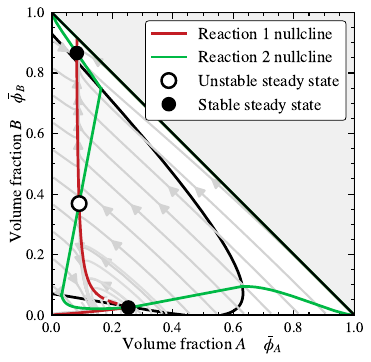}\label{fig:2b}}}
     \caption{
     \textbf{Full phase diagram of excitable systems:} 
     The full phase space, instead of the subset displayed in Fig.~\ref{fig:3} and Fig.~\ref{fig:3_appendix}, shown in (a) and (b) respectively, shows that there are two stable steady state points. Different initial conditions lead to different attractors, roughly separated by the total volume fraction $\psi$ of the unstable steady state point.
}
\label{fig:appendix3}
\end{figure}

The phase diagrams in Fig.~\ref{fig:2}(a), Fig.~\ref{fig:3}, and Fig.~\ref{fig:3_appendix} only show a subset of the whole phase space. Therefore, the entire phase diagrams are shown in Fig.~\ref{fig:appendix2} and Fig.~\ref{fig:appendix3}. For the former, we see how the unfueled case (blue) selects a specific tie-line, while the fueled nullclines cross different tie-lines with different phase compositions. Furthermore, the full binodal makes it apparent how $A$-$C$ only barely phase separates, while $B$-$C$ is strongly repulsive. These parameters are chosen to achieve the conditions for bistability, as discussed in Appendix \ref{sec:analytics}, by making the angle $\alpha$ in Eq.~\eqref{eq:crit_interactions} large.\par

The full phase space for an excitable droplet, as depicted in Fig.~\ref{fig:3} and Fig.~\ref{fig:3_appendix}, is shown in Fig.~\ref{fig:appendix3}(a,b). Here we see that the system is a bistable excitable medium, as there are two stable state points (black dots), and one unstable steady state (white dot). Even though there are two stable steady state points, only the lower one is excitable, as the flow lines show. The separatrix between the two equilibria is approximately given by the conserved line $\bar{\psi}$ of the unstable steady state. All initial conditions with $\bar{\psi}$ smaller than this value will create an excitable system.

\section{Effective description of bistable active droplets}\label{sec:effective_bistab}
For the single reaction $r_1$, defined in Eq.~\eqref{eq:r_1_pathway}, the evolution of the system is efficiently described by the dynamics of two fields; the reaction extent $\bar{\xi}\equiv(\bar{\phi}_B - \bar{\phi}_A)/2$ and the total volume fraction $\bar{\psi}\equiv(\bar{\phi}_A - \bar{\phi}_B)/2$. With this choice of coordinates, $\bar{\psi}$ is a constant $\bar{\psi}_0$, which is set by the initial conditions, while the change in $\bar{\xi}$ is equal to the phase-averaged reaction rate
\begin{equation}
    \partial_t\bar{\xi} = \bar{r}_1(\bar{\xi},\bar{\psi}_0)\,. \label{eq:xi}
\end{equation}
Here, we will expand the reaction rate at two different $\xi$-values, one where the system is homogeneous and one where it is phase-separated.  By connecting the two cases, we find a piece-wise function that captures the normal form of bistable droplets given in Eq.~\eqref{eq:effective_equations}, and displayed in Fig. \ref{fig:2}(b).

Let us start with the homogeneous case. The fluxes around the homogeneous chemical equilibrium point, denoted $(\bar{\xi}_0, \bar{\psi}_0)$, is approximately linear,
\begin{equation}
   \partial_t\bar{\xi} = -\left(\bar{\xi} - \bar{\xi}_0\right)h_1(\bar{\xi}_0,\bar{\psi}_0)\,, \label{eq:approx1}
\end{equation}
where $h_1>0$ makes  $\bar{\xi}_0$ a stable steady state. The first order coefficient of the Taylor expansion of $\bar{r}_1$, denoted $h_1$, takes the form
\begin{equation}
    h_1 = r_1^\rightharpoonup\left[\frac{2\bar{\psi}_0}{\bar{\psi}_0^2 - \bar{\xi}_0^2} - \pdv{}{\bar{\xi}}\ln{\left(\frac{\gamma_A}{\gamma_B} \right)}\Bigg\vert_{\bar{\xi}_0,\bar{\psi}_0}\right]\,, \label{eq:h_1}
\end{equation}
where we have decomposed the chemical potential in terms of the reference chemical potentials $\mu_i^0$ and the activity coefficients $\gamma_i$, as defined through Eq.~\eqref{eq:decomp_chempot}.
The chemical equilibrium point $(\bar{\xi}_0,\bar{\psi}_0)$ around which we have expanded, defined by $\mu_A=\mu_B$, is given by
\begin{equation}
    \frac{\bar{\xi}_0}{\bar{\psi}_0} = 1- \frac{2}{1+\frac{\gamma_A}{\gamma_B}\exp{\frac{\mu_A^0 + \mu_B^0}{k_BT}}}\,.
\end{equation}
The chemical forward and backward rates as used in Eq.~\eqref{eq:h_1}, $r^\rightharpoonup_1$ and $r^\leftharpoondown_1$ respectively, are by definition positive
\begin{equation}
    r_1^\rightharpoonup\equiv k_1\exp{\frac{\mu_A}{k_BT}}\, ,\qquad r_1^\leftharpoondown \equiv k_1\exp{\frac{\mu_B}{k_BT}}\,. \label{eq:r_f_or_r_back}
\end{equation}
Furthermore, the first term in the square bracket of Eq. \eqref{eq:h_1} is by the definitions of $\bar{\psi}$ and $\bar{\xi}$ positive, while the second term can take both signs. Using our form of the chemical potential (Eq.~\eqref{eq:chem_pot}), the full expression simplifies to
\begin{equation}
    h_1 = r_1^\rightharpoonup\left[\frac{2\bar{\psi}_0}{\bar{\psi}_0^2 - \bar{\xi}_0^2} - 2\frac{\chi_{AB}}{k_BT}\right]\,, \label{eq:h_1_def}
\end{equation}
independently of the number of components.

We now perform a Taylor expansion of $\bar{r}_1$ for the case where the system is phase-separated. For phase-separated systems, changes in $\bar{\xi}$ now result in changes in the phase volumes and the composition of the phases. The local maxima in the phase-separated domain in Fig.~\ref{fig:2}(a), denoted as $(\bar{\xi}_2, \bar{\psi}_0)$, can be approximated as
\begin{equation}
    \partial_t\bar{\xi} = h_0 - \left(\bar{\xi} - \bar{\xi}_2\right)^2 \, h_2(\bar{\xi}_2,\bar{\psi}_0)\,, \label{eq:approx2}
\end{equation}
where $h_0$ is a constant given by $h_0=\bar{r}_1(\bar{\xi}_2,\bar{\psi}_0)$, and $h_2$ is the negative of the second order coefficient of the Taylor expansion, i.e., $h_2=-\partial_{\bar{\xi}}^2\bar{r_1}\big\vert_{\bar{\xi}_2,\bar{\psi}_0}/2$, is given by
\begin{align}
    2h_2 = -& \pdv[2]{}{\bar{\xi}}\left(r_1^\rightharpoonup - r_1^\leftharpoondown\right) \nonumber \\ &- \left[\exp{\frac{\delta}{k_BT}}-1\right]\pdv[2]{}{\bar{\xi}}\left(\frac{V^\RN{2}r_1^\rightharpoonup}{V}\right)\,. \label{eq:h_2}
\end{align}
Notably, the composition in each phase ($\phi_i^{\RN{1}/\RN{2}}$) have a non-linear transcendental dependence on the phase-averaged quantities ($\bar{\xi},\bar{\psi}$) which must be known to perform the derivatives in Eq. \eqref{eq:h_2}.
Further note that the conditions of phase coexistence (Eq.~\eqref{eq:phase_coex}) allow us to omit the phase index of the forward and backward reaction rates (defined in Eq.~\eqref{eq:r_f_or_r_back}). All non-equilibrium effects are captured by the second line of Eq.~\eqref{eq:h_2}, where volume changes are irrelevant for equilibrium systems ($\delta=0$). \par 
For bistability to occur $h_0$, as defined in Eq.~\eqref{eq:r_bar_rewrite}, must be positive. Remember that both the forward and backward reaction rates are positive by definition (Eq.~\eqref{eq:r_f_or_r_back}). 
For a sufficiently strong fuel strength $\delta$, the coefficient $h_0$ can become positive, as we see in Eq. \eqref{eq:r_bar_rewrite}.

By ensuring continuity of $\partial_t\bar{\xi}$ at the binodal $\bar{\xi}=\xi^{\RN{1}}(\bar{\psi}_0)$, we can find an expression for $h_2$ in terms of $h_0$ and $h_1$, yielding
\begin{equation}
    h_2 = \frac{h_1\left( \xi^\RN{1} - \xi_0 \right) + h_0}{\left(\xi^\RN{1}-\bar{\xi_2}\right)^2}\,.\label{eq:approx3}
\end{equation}
This alters the curvature at the expansion point $(\bar{\xi}_2, \bar{\psi}_2)$ while keeping the constant offset fixed. The choice of expressing $h_2$ in terms of the other coefficients is done to reduce the complexity of the normal form, while still capturing the essential role of the sign of $h_0$.

From the two expressions, we can write a piece-wise function for $\partial_t\bar{\xi}$ as given in the main text in Eq.~\eqref{eq:effective_equations}. Usually, the normal form of bistable chemical reactions in homogeneous systems originates from a cubic relationship between the reaction extent and its time derivative~\cite{epsteinNonlinearChemicalDynamics1996,novakDesignPrinciplesBiochemical2008}. In contrast, for phase-separated systems the minimal prerequisite of bistability is a quadratic relationship between $\xi$ and its time derivative. By utilizing the different behavior of homogeneous and phase-separated systems, bistability can occur for simpler reaction schemes. In the following Appendix, we perform the same expansion for systems where $\bar{\psi}$ is dynamic.

\section{Effective description of excited active droplets}\label{sec:effective_active}

We follow a similar procedure as in Appendix~\ref{sec:effective_bistab} to describe an excitable system. The key difference is a further chemical reaction with the rate $r_2$, defined in Eq.~\eqref{eq:r_2_pathway}, which makes $\bar{\psi}$ change in time according to
\begin{equation}
    \partial_t\bar{\psi}= \bar{r}_2\, ,
\end{equation}
while the kinetic equation of $\bar{\xi}$ in Eq.~\eqref{eq:xi} remains unchanged. The procedure of using piece-wise functions by treating the homogeneous and phase-separated cases separately will still be followed. We begin with the homogeneous case. 

As the additional reaction $\bar{r}_2$ obeys detailed balance with phase-independent rate coefficients $(k_2^\RN{1}=k_2^\RN{2})$, the reaction rate $r_2$ is equal in both phases. This makes changes in phase volumes not affect the phase-averaged reaction rates, such that the change in phase-averaged reaction rate $\partial_{\bar{\psi}}\bar{r}_2$ is equal to the reaction rate change in either phase $\partial_{\bar{\psi}}\bar{r}_2=\partial_{\bar{\psi}}r^\RN{1}_2=\partial_{\bar{\psi}}r^\RN{2}_2$. As a consequence, its expansion can approximately be made independently of phase separation, where we find 
\begin{equation}
    \partial_t\bar{\psi} = -\left(\bar{\xi}-\bar{\xi}_1\right)f_\xi\left(\bar{\psi}_1, \bar{\xi}_1\right) - \left(\bar{\psi} - \bar{\psi}_1\right)f_\psi(\bar{\psi}_1, \bar{\xi}_1)\, .
\end{equation}
The expansion point $(\bar{\psi}_1, \bar{\xi}_1)$ is chosen as the unstable steady state point in Fig.~\ref{fig:4}(a), allowing us to set the zeroth order contribution to zero. Furthermore, the expansion is stable to changes in $\bar{\psi}$ ($f_\psi>0$),
where we have 
\begin{align}
    \frac{f_{\psi}}{2} &= \bar{r}_2^\rightharpoonup\left( \frac{2\bar{\xi}_1^2-\bar{\psi}_1}{\left( \bar{\psi}_1^2 - \bar{\xi}_1^2 \right)\left(2\bar{\psi}_1-1\right)} + \frac{\chi_{AB}-2\chi_{BC}-2\chi_{AC}}{k_BT}\,\right)\,,\nonumber \\
    \frac{f_{\xi}}{2} &= \bar{r}_2^\rightharpoonup\left( \frac{\bar{\xi}_1}{ \bar{\xi}_1^2 - \bar{\psi}_1^2} + \frac{\chi_{AC}-\chi_{BC}}{k_BT}\,\right)\,, \label{eq:def_rhos}
\end{align}
where the first term of $f_\psi$ is always positive. The sign of $f_\xi$ can be both positive and negative. Performing a Taylor expansion of $\bar{r}_1$ now requires an expansion in both variables $\bar{\xi}$ and $\bar{\psi}$. The homogeneous expansion becomes
\begin{equation}
    \partial_t\bar{\xi} = -\left(\bar{\xi} - \bar{\xi}_0\right)h_\xi - \left(\bar{\psi} - \bar{\psi}_0\right)h_\psi\,. \label{eq:approx4}
\end{equation}
Here, $h_\xi$ takes the same form as in Eq. \eqref{eq:h_1_def}, and $h_\psi$ can take both signs 
\begin{equation}
    h_\psi = 2r_1^\rightharpoonup\left( \frac{-\bar{\xi}_0}{{\bar{\psi}^2_0 - \bar{\xi}^2_0}} + \frac{\chi_{AC}-\chi_{BC}}{k_BT}\right) \, . \label{eq:h_psi}
\end{equation}
For the phase-separated expansion at $(\bar{\xi}_2, \bar{\psi}_1)$, the only addition in comparison to the derivation in Appendix~\ref{sec:effective_bistab}, which granted the normal form in Eq.~\eqref{eq:effective_equations}, is the expansion in $\bar{\psi}$, which yields a first-order coefficient 
\begin{equation}
    \tilde{h}_\psi = \pdv{}{\bar{\psi}}\left(r_1^\leftharpoondown - r_1^\rightharpoonup\right)  \, . \label{eq:h_psi_tilde}
\end{equation}
Such that the normal form for $\bar{\xi}$ becomes
\begin{equation}
    \partial_t\bar{\xi} = 
\begin{cases}
   -\left(\bar{\psi} - \bar{\psi}_0\right)h_\psi -  \left(\bar{\xi} - \bar{\xi}_0\right)h_\xi,\qquad\quad\, \bar{\xi} \leq \xi^\RN{1},\\
   \bar{r}_1 - \left(\bar{\psi} - \bar{\psi}_1\right)\tilde{h}_\psi  - h_{\xi\xi}\left(\bar{\xi} - \bar{\xi}_2\right)^2,\quad \bar{\xi} > \xi^\RN{1}\,.
\end{cases}
\end{equation}
The system follows the two coupled equations for the phase-averaged total volume fraction $\bar{\psi}$ and for the phase-averaged reaction extent $\bar{\xi}$ given in Eq.~\eqref{eq:effective_active}. The coefficients $f_{\xi}$, $h_\psi$, and $\tilde{h}_\psi$ induces a bilateral coupling between $\bar{\psi}$ and $\bar{\xi}$. Similar to other oscillating chemical reactions~\cite{epsteinNonlinearChemicalDynamics1996,novakDesignPrinciplesBiochemical2008}, the kinetic equation for $\bar{\xi}$  is a sum of a coupling between the two quantities and a bistability. The combination of bistability and a negative feedback between $\bar{\psi}$ and $\bar{\xi}$ induces self-sustained stable chemical oscillations. The negative feedback emerges when large $\bar{\psi}$-values $(\bar{\psi}>\bar{\psi_1})$ leads to a decrease in $\bar{\xi}$ through $\tilde{h}_{\psi}>0$, while small $\bar{\psi}$-values $(\bar{\psi}<\bar{\psi_0})$ leads to an increase in $\bar{\xi}$ through $h_{\psi}<0$. When this is combined with $\bar{\xi}>\bar{\xi}_1$ decreasing $\bar{\psi}$ and $\bar{\xi}<\bar{\xi}_1$ increasing it, i.e. $f_\xi>0$, self-sustained cycles in $\bar{\xi}$ and $\bar{\psi}$ appear. Note that the essential property of $r_2$ is to make the otherwise conserved quantity $\bar{\psi}$ dynamic. The same normal form might therefore be achievable when allowing material exchange with a reservoir \cite{haugerudNonequilibriumWetDry2024a}.\par

\section{Excitability to an active droplet}\label{sec:excited_homo}

\begin{figure}
    \centering
      \makebox[\textwidth][c]{
      \includegraphics[width=\textwidth]{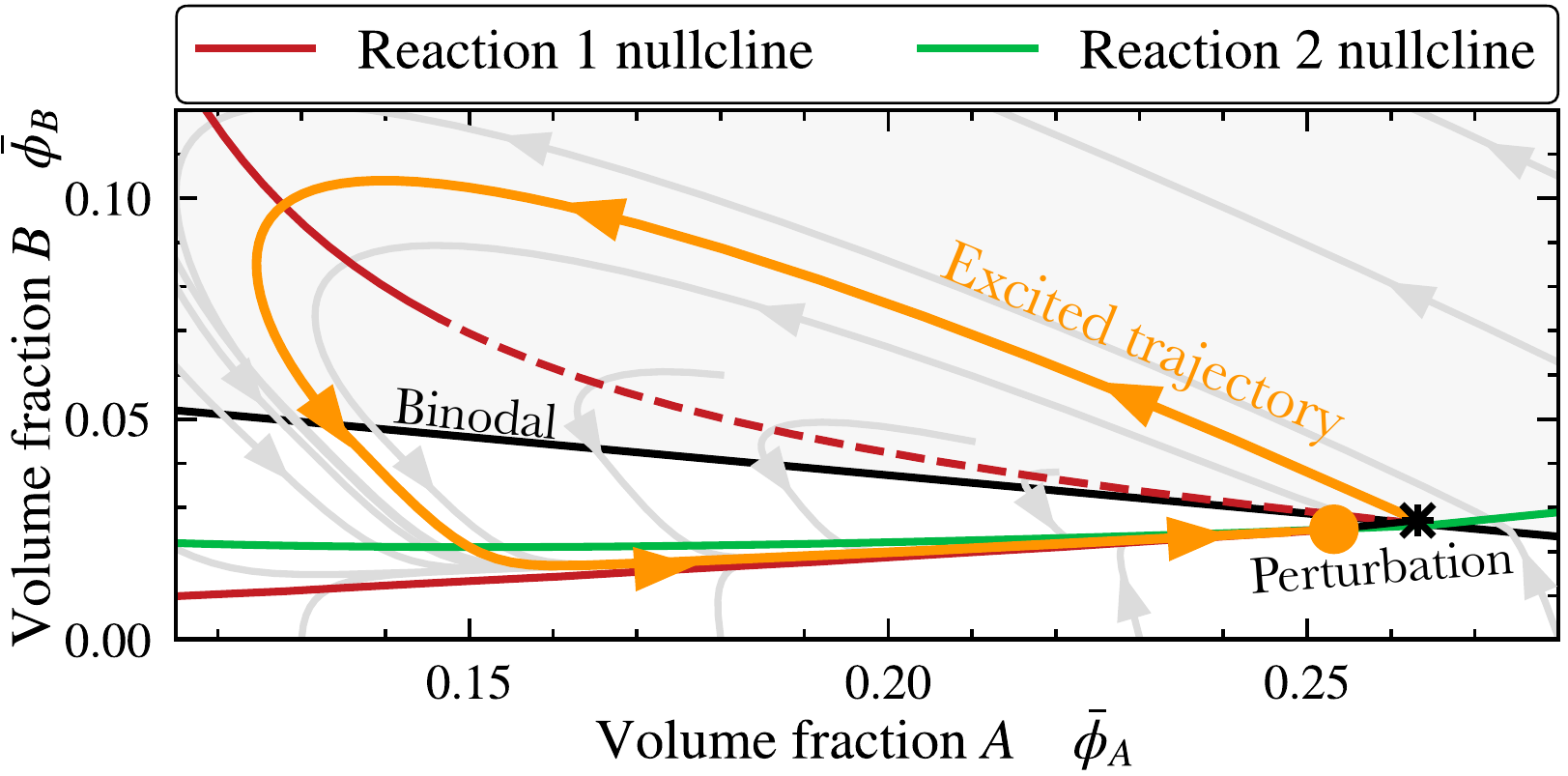}}
     \caption{
     \textbf{Excitability to an active droplet:}
     When considering two chemical reactions (reaction $r_1$ Eq.~\eqref{eq:r_1_pathway} and reaction $r_2$ Eq.~\eqref{eq:r_2_pathway}),
the active droplet system becomes excitable.    
     For NESS close to the binodal, a compositional perturbation beyond the binodal line (black cross) can excite the system, forming an active droplet. As a result, the system follows an extended trajectory (orange)  with  changes in average composition far beyond the initial  perturbation. 
     During this excited trajectory, the system has a refractory period in which it is no longer excitable. It relaxes back to the NESS, where the reaction nullclines for both reactions intersect (green and red lines).   
     }
     \label{fig:3_appendix}
\end{figure}

Similar to the excitable droplet displayed in Fig. \ref{fig:4}, a homogeneous system with the two reactions $r_1$ and $r_2$ can become excitable upon a droplet forming perturbation. To achieve this, the intersection of the two nullclines must occur outside the binodal. Thus, a droplet forming perturbation into the binodal results in an excited trajectory much longer than the perturbation, as seen in Fig. \ref{fig:3_appendix}. The dilute branch of the binodal corresponds to an excitation threshold.  During the excited trajectory, the droplet volume and composition change until it dissolves and returns to its homogeneous steady state. The system is not excitable along the trajectory, thus exhibiting a refractory period.

\section{Finding the critical fuel strength $\delta^*$}\label{sec:analytics}

Here, we derive an approximate expression for the critical fuel strength $\delta^*$, above which bistability occurs. To this end, we use the expression for the volume of the dense phase $V^\RN{2}$ at the phase-separated nullcline given in Eq.~\eqref{eq:NESS}, and insert it into the definition of the phase-averaged total volume fraction $\bar{\psi}$ using Eq.~\eqref{eq:tot_vol_frac}. Finding the conditions for bistability is equivalent to finding non-equilibrium steady states with $\bar{\psi} < \psi_0^\RN{1}$, where $\psi_0^\RN{1} = \phi_A^0 + \phi_B^0$ is the total volume fraction where the dilute branch of the binodal intersects the nullcline of $r_1$ (Eq. \eqref{eq:r_1_pathway}). Solving for $\delta^*$ gives an expression the critical fuel strength necessary for bistability as a function of $\Delta\psi=\bar{\psi}-\psi^\RN{1}$:
\begin{equation}
    e^{\frac{\delta^*}{k_BT}} = 1 + (1-e^{\frac{-\Delta\mu}{k_BT}})\left(\frac{k^\RN{1}_1}{k^\RN{2}}\left(1 + \frac{\psi^\RN{2} - \psi^\RN{1}}{\bar{\psi} - \psi^\RN{1}} \right) - 1\right)\, , \label{eq:analytic_first_exp}
\end{equation}
where $\Delta\mu=\mu_A-\mu_B$ is the chemical potential difference between $A$ and $B$. As the fuel converts $A$ to $B$, the chemical potential difference is negative; $\Delta\mu<0.$ So far, no approximations have been made.

\begin{figure}
    \centering
      \makebox[\textwidth][c]{
      \includegraphics[width=0.75\textwidth]{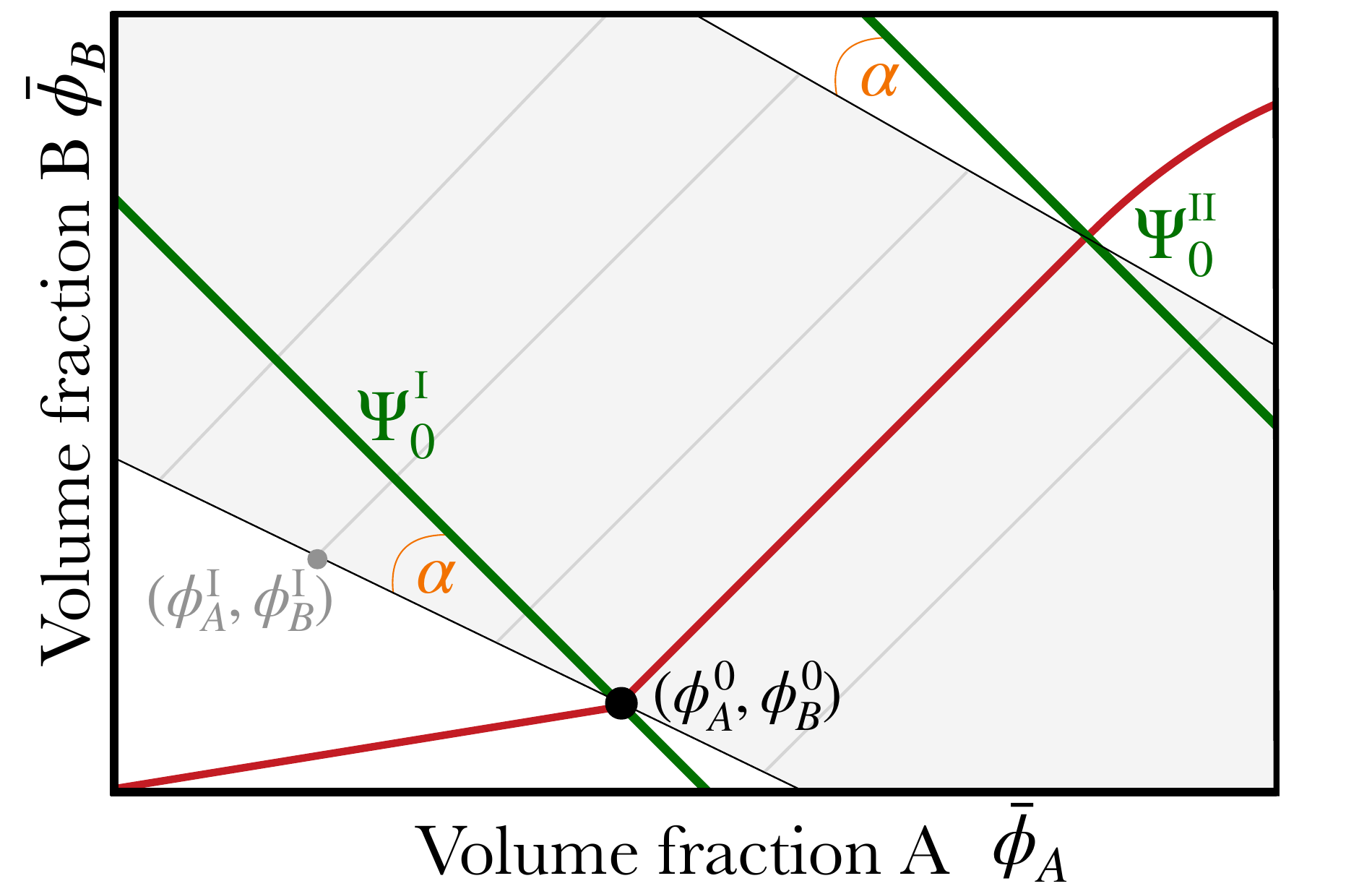}}
     \caption{
     \textbf{Approximate phase diagram:} To approximate the chemical potential around the chemical equilibrium point of the dilute binodal branch $(\phi_A^0, \phi_B^0)$, we approximate the binodal as a straight line with angle $\alpha$ relative to the conserved line $\psi_0^\RN{1}=\phi_A^0+\phi_B^0$. With this approximation, a point align the binodal can be written as $(\phi_A^0 - \epsilon\sin{\left(\alpha + \pi/4\right)}, \phi_B^0 - \epsilon\sin{\left(\alpha - \pi/4\right)})$, leading to the expression for the chemical potential in Eq.~\eqref{eq:approx_chem}.}
     \label{fig:appendix_analytics}
\end{figure}

Due to the non-linearity of Eq.~\eqref{eq:analytic_first_exp}, we perform a Taylor-expansion at the point where the $r_1$ nullcline intersects the dilute branch of the binodal, which we denote $(\phi_A^0,\,\phi_B^0)$. We investigate small deviations around this point, i.e., close to chemical equilibrium where $\abs{\Delta\mu}\ll k_BT$. Furthermore, we follow Ref.~\cite{bauermannCriticalTransitionIntensive2024} and approximate the tie-lines to be orthogonal to the binodal. This implies that $(\psi^\RN{2}-\psi^\RN{1})$ is a constant along the binodal and can be set equal to the difference at the chemical equilibrium tie-line $(\psi^\RN{2}_0-\psi^\RN{1}_0)$. With these assumptions, we find
\begin{align}
    \begin{split}
    e^{\frac{\delta^*}{k_BT}} = 1 - \frac{\Delta\mu}{k_BT}\left[\frac{k^\RN{1}_1}{k^\RN{2}_1}\left(1+\frac{\psi_0^\RN{2} - \psi_0^\RN{1}}{\Delta\psi}\right) - 1\right]& \label{eq:early_delta_star} \\ + \frac{1}{2}\left(\frac{\Delta\mu}{k_BT}\right)^2\left[\frac{k^\RN{1}_1}{k^\RN{2}_1}\left( 1 + \frac{\psi_0^\RN{2} - \psi_0^\RN{1}}{\Delta\psi} \right)-1\right]& \, , \end{split}
\end{align}
where the Taylor expansion is truncated above second order, ignoring $\mathcal{O}(\Delta\mu/k_BT)^3$. We continue by introducing the angle $\alpha$ of the dilute branch of the binodal at $(\phi_A^0,\,\phi_B^0)$, relative to the conserved line of $\psi^\RN{1}_0 = \phi_A^0 + \phi_B^0$ at the same point. This definition of the angle is such that $\alpha=0$ gives a binodal parallel to the conserved line, where the dilute branch of the binodal can be approximated as $(\phi_A^0 - \epsilon\sin{\left(\alpha + \pi/4\right)}, \phi_B^0 - \epsilon\sin{\left(\alpha - \pi/4\right)})$, where $\epsilon \simeq \Delta\psi/(\sqrt{2}\sin{\left(\alpha\right)})$. An illustration is given in Fig.~\ref{fig:appendix_analytics}. Performing this Taylor expansion in the chemical potential difference, we find $\Delta\mu\approx\Delta\psi\,\theta$, where
\begin{align}
    \theta &= \frac{\cos{\alpha}\left(\phi_A^0\right)^{-1} +  \sin{\alpha}\left(\phi_B^0\right)^{-1}}{\sqrt{2}\sin{\alpha}} \label{eq:approx_chem}\\ &+ \frac{\chi_{BC}-\chi_{AC}}{k_BT} + \frac{1}{\tan{\alpha}}\frac{\chi_{AB}}{k_BT}\, .\nonumber
\end{align}
Using this approximate expression for the chemical potential in Eq.~\eqref{eq:early_delta_star}, we find
\begin{align}
    \exp{\frac{\delta^*}{k_BT}} &= 1 + \frac{k^\RN{1}_1}{k^\RN{2}_1}\theta\left(\psi_0^\RN{2} - \psi_0^\RN{1}\right) \label{eq:full_delta*} \\
    &- \Delta\psi\,\theta\left(1 - \frac{k^\RN{1}_1}{k^\RN{2}_1}\left( 1 + \theta\frac{\psi^\RN{2}_0-\psi^\RN{1}_0}{2}\right) \right) \nonumber \\
    &+ \left(\Delta\psi\right)^2\frac{\theta^2}{2}\left(1-\frac{k^\RN{1}_1}{k^\RN{2}_1}\right)\nonumber\,.
\end{align}
The terms independent of $\Delta\psi$ give the expression for the minimal fuel strength needed for bifurcation, as shown in Eq.~\eqref{eq:ana_deltastar} and Fig.~\ref{fig:2}(e), where we have separated the thermodynamic parameters into $\Theta\equiv \left( \psi_0^\RN{2}-\psi_0^\RN{1}\right)\theta$, such that
\begin{align}
    \Theta &= \left(\psi_0^\RN{2} - \psi_0^\RN{1}\right)\Big(\frac{\chi_{BC}-\chi_{AC}}{k_BT} +  \frac{1}{\tan{\alpha}}\frac{\chi_{AB}}{k_BT} \nonumber \\
    &+\frac{\cos{\alpha}\left(\phi_A^0\right)^{-1} +  \sin{\alpha}\left(\phi_B^0\right)^{-1}}{\sqrt{2}\sin{\alpha}}\Big)\, .\label{eq:crit_interactions}
\end{align}
Notably, the expression diverges at an angle $\alpha=0$, and any angle $\alpha < 0$ gives a large negative value of $\Theta$, which does not yield physical solutions of $\delta^*$ in Eq.~\eqref{eq:ana_deltastar}. This can be understood geometrically, as the $r_1$ nullcline has to lie within the binodal for multiple intersections with $\psi$ to occur, and therefore, $\psi$ has to also lie within the binodal. Note that $\alpha$ is in general negative when $\chi_{AC}>\chi_{BC}$, and positive when $\chi_{AC}<\chi_{BC}$, making the latter a necessity for bistability. 
The minimal value of $\Theta$, which occurs for $\alpha=\pi/2$, is finite. Consequently, the critical fuel strength $\delta^*$ never approaches zero, and a finite fuel strength is always necessary for bistability. Note further that large values of $\alpha$, which lowers the critical fuel strength, occurs when $\chi_{BC}\gg\chi_{AC}$, thereby increasing the constant offset of $\Theta$. It is, therefore, not trivial to minimize the expression for $\Theta$. The predicted critical fuel strength is compared with the analytic expression in Fig.~\ref{fig:2}(e), which agrees with the numerical solution within $10\%$. As $\Delta\psi$ increases the deviation from chemical equilibrium increases and higher order terms in $\Delta\mu$ are need for an agreement, creating a growing discrepency between the analytic and numerical result. 

For the case where the reaction rate coefficient is equal in both phases, the expression simplifies to
\begin{equation}
    e^{\frac{\delta^*}{k_BT}} = 1 - \left(\psi^\RN{2}_0 -\psi^\RN{1}_0 \right)\theta + \Delta\psi \left(\psi^\RN{2}_0 -\psi^\RN{1}_0 \right) \frac{\theta^2}{2}\, ,
\end{equation}
as the second order contribution of $\Delta\psi$ in Eq.~\eqref{eq:full_delta*} cancels. 

\section{Oscillations in droplet volume}\label{sec:volume_oscillations}

\begin{figure}
    \centering
      \makebox[\textwidth][c]{
      \includegraphics[width=0.9\textwidth]{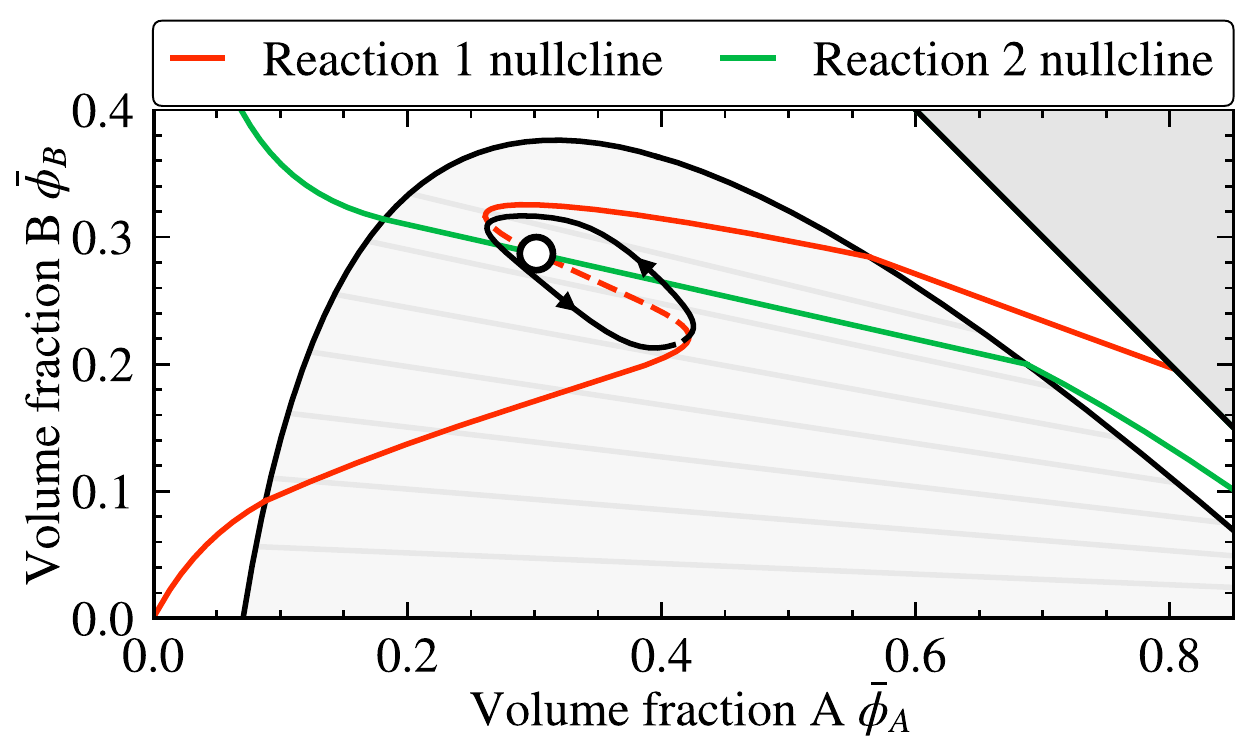}}
     \caption{
     \textbf{Oscillations in droplets volumes:} The same chemical reaction network can produce oscillations in the droplet volume. These oscillations are achieved by changing the interaction strengths between the components, altering the binodal and, therefore, the curvature of the reaction nullcline. }
     \label{fig:appendix_intermediateDrops}
\end{figure}

We have investigated the case of a system where both $A$ and $B$ independently phase-separate from the solvent. The same reaction network can produce fundamentally different oscillations by changing the interactions between the components. For example, it could make large droplet volumes unstable, though this is similar to the system we studied here. \par 
An interesting case is a system where the reaction nullcline becomes unstable in the center of the binodal instead of by either the dilute or dense branch, as depicted in Fig.~\ref{fig:appendix_intermediateDrops}. Such a system does not result in formation and dissolution cycles but droplet volume oscillations. The oscillations emerge from an interval of droplet compositions being unstable. As the droplet grows and changes composition, it becomes unstable, leading to a fast change in composition and a decrease in droplet volume. This fast composition change crosses the nullcine of $r_2$, leading to a turnover from $C$ to $A$ and $B$. Again, this turnover leads to an unstable droplet composition where diffusive and reactive fluxes cannot balance, leading to a fast increase in droplet volume, continuing indefinitely.

The discussed oscillation type originates from phase diagrams where the dilute phase can become dense enough for the fuel to partition into the dilute and dense phases. This reduces the effect of the fuel, as it affects reactions similarly in both phases. In the limiting case of the fuel affecting the reaction equally in both phases, the system is mappable to an equilibrium system with a change in internal energy $\tilde{\omega}_A = \omega_A + \delta$. For the case in Fig.~\ref{fig:appendix_intermediateDrops}, the fueling threshold is $\psi_{1/2}=0.45$, with a less steep fuel transition width than previously, $\kappa=0.04$. 

\end{document}